\documentclass[
	aps, pra,
	superscriptaddress, twocolumn,
	floatfix,
	10pt
]{revtex4-2}

\usepackage{silence}
\usepackage{parskip}
\usepackage{soul}

\usepackage[final]{graphicx}
\graphicspath{{./figures/}}
\usepackage{times,bbm,amsmath,amssymb}
\usepackage{mathtools}
\usepackage{microtype}
\usepackage{epsfig,color}
\usepackage[dvipsnames]{xcolor}
\usepackage{hyperref}
\hypersetup{
    colorlinks = true
}
\usepackage{cleveref}

\usepackage{libertine}

\usepackage{float,siunitx}

\usepackage{acronym}
\usepackage{easyReview}

\usepackage{tikz}

\usepackage{standalone}

\usepackage{physics}

\usepackage{blindtext}

\newcommand{\normord}[1]{:\mathrel{#1}:}
\newcommand{\calT}{{\mathcal{T}}}

\newcommand{\calM}{{\mathcal{M}}}

\def\<#1>{\mathinner{\langle#1\rangle}}

\begin{document}

\title{Experimental certification of level dynamics in single-photon emitters}

\author{Luk\' a\v s Lachman}
\thanks{equal contribution}
\affiliation{Department of Optics, Palack{\'y} University, 17. Listopadu 12, 771 46 Olomouc, Czech Republic}
\author{Ilya P. Radko}
\thanks{equal contribution}
\affiliation{Center for Macroscopic Quantum States (bigQ), Department of Physics, Technical University of Denmark, 2800 Lyngby, Denmark}
\author{Maxime Bergamin}
\affiliation{Center for Macroscopic Quantum States (bigQ), Department of Physics, Technical University of Denmark, 2800 Lyngby, Denmark}
\author{Ulrik L. Andersen}
\affiliation{Center for Macroscopic Quantum States (bigQ), Department of Physics, Technical University of Denmark, 2800 Lyngby, Denmark}
\author{Alexander Huck}
\affiliation{Center for Macroscopic Quantum States (bigQ), Department of Physics, Technical University of Denmark, 2800 Lyngby, Denmark}
\author{Radim Filip}
\affiliation{Department of Optics, Palack{\'y} University, 17. Listopadu 12, 771 46 Olomouc, Czech Republic}

\begin{abstract}
Emitters of single-photons are essential resources for emerging quantum technologies and developed within different platforms including nonlinear optics, atomic and solid-state systems. The energy level structures of emission processes are critical for reaching and controlling high-quality sources. The most commonly applied test uses a Hanbury-Brown and Twiss (HBT) setup to determine the emitter energy level structure based on fitting temporal correlations of photon detection events. However, only partial information about the emission process is extracted from such detection, that might be followed by an inconclusive fitting of the data. This process predetermines our limited ability to quantify and understand the dynamics in the photon emission process that are of importance for the applications in communication, sensing and computing. In this work, we present a complete analysis based on all normalized coincidences between detection and no-detection events recorded in the same HBT setup to certify expected properties of an emitted photonic state. As a proof of concept we apply our methodology to single nitrogen-vacancy centers in diamond, in which case the certification conclusively rejects a model based on a two-level emitter that radiates a photonic states mixed with any classical noise background.
\end{abstract}

\maketitle



\section{Introduction}

Quantum technologies with photons, atoms, solid-state and remote superconducting systems rely on the structure, quality and development of photonic light sources. Efficient emitters of pure single photons are the primary targets of this ongoing rally~\cite{Eisaman2011,Aharonovich2016}, and their development is the first milestone towards the realization of more advanced photonic sources~\cite{Huber2017,Khoshnegar2017,Oestfeldt2022}. Multilevel structures of emitters require novel excitation schemes to control and improve the emission of single-photons~\cite{Thomas2021,Knall2022,Wei2022}.
The Hanbury-Brown and Twiss (HBT)~\cite{Brown1954} setup utilizing a beam splitter and a pair of single-photon sensitive detectors is the traditional measurement tool for single-photon emitters, applied to understand and tailor the emission mechanism and for quantifying the performance of the source. Normalized coincidences of detection events between both detectors with controllable delay yield information about the source properties including photon bunching and anti-bunching~\cite{Kimble1977,Grangier1986}. In the limit of weak optical emission, these correspond to normalized Glauber’s second-order photon-photon correlation function~\cite{Sperling2012}. Such normalized coincidence detection events are independent of losses but sensitive to independent Poissonian background noise, the two main limitations of many experimental platforms. Fits to the second-order correlation function with respect to time delay between photodetection events can be used as a tool to investigate energy level dynamics and relaxation processes in the emitter~\cite{Kitson1998,Khramtsov2017} and to verify the presence of single-photon emitters~\cite{Shcherbina2014}. This traditional approach has, however, some intrinsic limitations making it insufficient for some technological applications. First, more than one model can in principle be used to fit the data, which can lead to an inconclusive interpretation of the measurement. A conclusive approach broadly used in quantum optics \cite{Mandel1995} relies on certification that a well-defined set of states does not suffice to explain a result obtained in a measurement. 
Second, correlation functions do not find use in some application areas. Employing photo-click probabilities in a temporal mode instead of correlation functions will directly enable the application in quantum sensing~\cite{Tan2019,Thekkadath2020} and quantum communication~\cite{Schimpf2021,Istrati2020,Konno2023}.
Third, information retrieved only from normalized coincidence detection is unavoidably biased as these quantities are insensitive to optical loss and very sensitive to independent noise. Another quantifier based on the no-click coincidences has thus been introduced forming a dual evaluation approach together with the second-order correlation function~\cite{Lachman2016}. 

In this article, we propose and demonstrate a threefold evaluation approach of single-photon emitters utilizing the full photo-click information available from a measurement with an HBT setup. We  apply the approach to a field test with nitrogen-vacancy centres in diamond, a well established source of single photons with a complex internal level structure~\cite{Doherty2013}. Following our advanced evaluation, we can conclusively certify that simple models do not explain the emitter structure and dynamics.

\begin{figure}[t!]
\centerline {\includegraphics[width=0.99\linewidth]{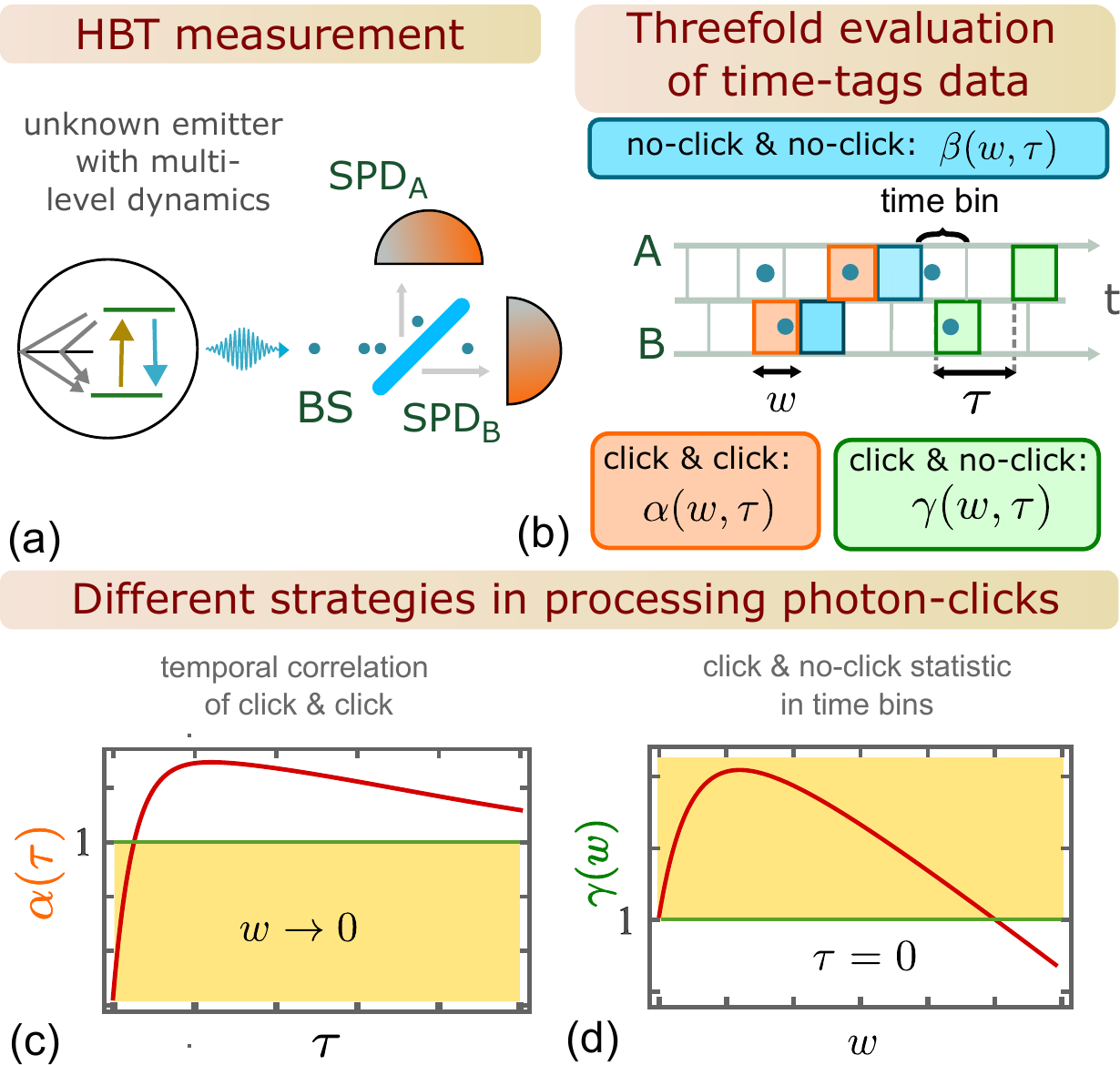}}
\caption{(\emph{a}) Measurement of light emitted from a source with unknown dynamics between its energy levels. The detection scheme consists of two single-photon sensitive detectors, SPD$_A$ and SPD$_B$, separated by a beam-splitter (BS) in the HBT configuration. (\emph{b}) Threefold evaluation of time-tagged data. The detectors convert the measured optical signal into electronically recorded time-tags depicted by blue dots as a function of time for both detectors.
The grid for both time axis A and B illustrates splitting into time bins of size $w>0$, while the temporal shift between these two grids suggests that we can generally compare events between time bins separated by a delay $\tau$.
The coloured fields illustrate click or no-click events that we employ for introducing the correlation functions $\alpha$, $\beta$ and $\gamma$ in Eq.~(\ref{defCorrF}) for given $w$ and $\tau$. (\emph{c})-(\emph{d}) Two approaches used for the analysis of dynamical aspects of a solid-state emitter. The traditional approach depicted in (\emph{c}) relies on the temporal correlation between click events evaluated by the correlation function $\alpha(\tau)$ for different delay times $\tau$ in the regime $w \rightarrow  0$. $\alpha(\tau)$ exhibits a 'dip' around $\tau=0$ signifying sub-Poissonian statistics (yellow region).  In the new approach shown in (\emph{d}), we choose $\tau=0$ and evaluate the click statistics in a time bin with of a given size $w$. The measured correlation function $\gamma(w)$ is characterized by its maximum $\bar{\gamma}(w)\equiv \max_w \gamma(w)$ over $w$. Reaching $\bar{\gamma}>1$ proves sub-Poissonian statistics (yellow region) of the emitted light. }
  \label{fig:2ls}
\end{figure}

\section{Threefold evaluation}
\label{sec_definitions}
The HBT configuration~\cite{Brown1954} comprises a beam splitter (BS) that divides incident light between two single-photon sensitive detectors (SPDs). Measurements on the light field that is not in the vacuum state give rise to "click" events at the output of each SPD and electronically recorded with time-tags (Fig.~\ref{fig:2ls}). 
Based on this detection, we evaluate a photonic state emitted to a well-defined temporal mode that we identify by its duration $w$. For a given $w$, we determine the detector response on the emitted light. We denote $P_1(w)$ as the probability of at least one click of a chosen SPD (irrespective of the response of the other SPD) in a time bin with a size $w$ and $P_{11}(w,\tau)$ as the probability of at least one coincidence click recorded by each SPDs in time bins of the same size $w$, but temporally shifted by~$\tau$~\cite{Grangier1986}.  
For the defined time bins, we also introduce $P_{00}(w,\tau)$ as the probability of no-click by both SPDs and the probability $P_{10}(w,\tau)$ of at least one click by one SPD and a no-click by the other (with a chosen SPD giving a click). Finally, we define $P_0(w)$ as the no-click probability of a SPD irrespective of the result of the other detector~\cite{Sekatski2022}. We extract all these probabilities from time-tagged data by determining the ratio between the number of time bins with a respective event (depending on a chosen $w$) and the total number of time bins. Note that unequal quantum efficiencies of SPDs or unbalancing of the beam-splitter will affect $P_1(w)$ and $P_0(w)$, and the analysis would depend on which SPD we employ for the measurement. Thus, we take the geometrical mean of the values obtained from one SPD and the other SPD to avoid this ambiguity. For $\tau=0$, we interpret the probability $P_{10}(w,\tau=0)$ as a probability of \emph{success}, $P_{00}(w,\tau=0)$ as a probability of \emph{failure}, and $P_{11}(w,\tau=0)$ as probability of \emph{error} according to an expected response of the HBT detection setup on single-photon states.

Although the probabilities of success $P_{10}(w,\tau)$, failure $P_{00}(w,\tau)$, and error $P_{11}(w,\tau)$ are mutually dependent, as shown in this paper, each of them reflects a different property of an incident photonic state. Consequently, an analysis focusing, for example, on $P_{11}(w,\tau)$ evaluates only one particular aspect of a single-photon source. On the contrary, considering together the probabilities of all possible outcomes enables an unbaised evaluation. A similar case based on multiple mutually dependent quantities is already well-known in photonics. Intensity fluctuations on a large signal can be simultaneously evaluated using the second-order correlation function~\cite{Glauber1963a}, the Fano-factor, the Mandel Q-factor \cite{Davidovich1996}, or the signal-to-noise ratio \cite{Mandel1979}. All these quantities depend on just the first two moments of the intensity. Here, a similar situation occurs while, however, our quantities are useful at the single-photon level and directly applicable for the HBT measurement. 

\begin{figure}[t!]
\centerline {\includegraphics[width=0.99\linewidth]{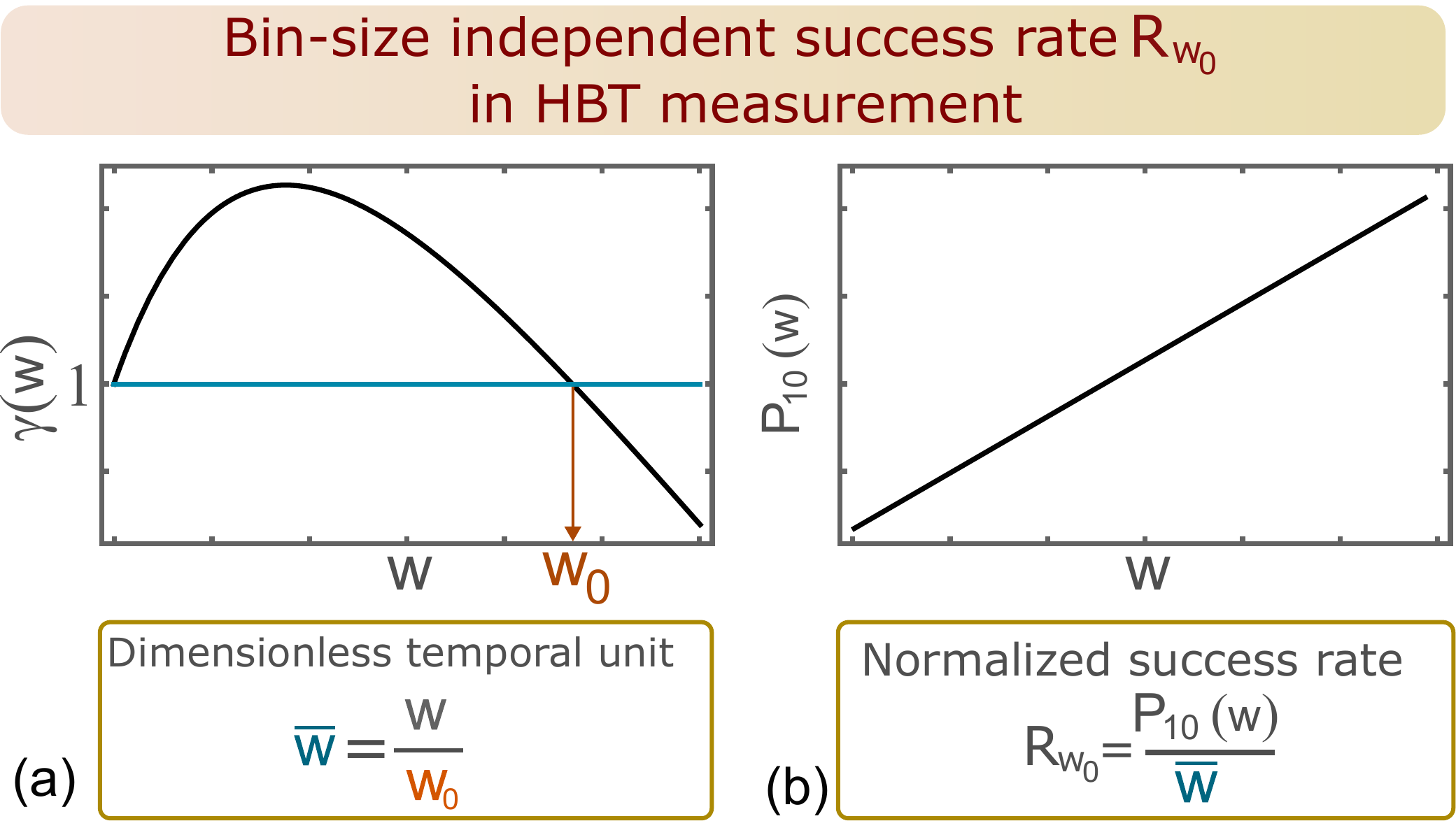}}
\caption{(a) Typical evolution of the correlation function $\gamma(w)$ with respect to the size of the time bin $w$. We determine $w_0$ such that $\gamma(w_0)=1$ and define the dimensionless time unit $\bar{w}\equiv w/w_0$. (b) Illustrative example for the evolution of the probability $P_{10}(w)$ as a function of $w$ in the regime with a low click probability. We introduce $R_{w_0}$ as $R_{w_0}\equiv P_{10}(w)/\bar{w}$. Since $P_{10}(w)$ grows linearly with $w$, $R_{w_0}$ is independent of the chosen $w$.}
  \label{fig:figR}
\end{figure}

For the analysis of photonic states, we define three normalized correlation functions $\alpha(w,\tau)$, $\beta(w,\tau)$ and $\gamma(w,\tau)$ as:
\begin{equation}
    \begin{aligned}
        \alpha(w,\tau) &=\frac{P_{11}(w,\tau)}{\left[P_{1}(w)\right]^2}\\
        \beta(w,\tau) &=\frac{P_{00}(w,\tau)}{\left[P_0(w)\right]^2}\\
        \gamma(w,\tau) &=\frac{P_{10}(w,\tau)}{P_0(w) P_1(w)},
    \end{aligned}
\label{defCorrF}
\end{equation}
which can be evaluated with respect to the variable time bin size $w>0$ or the time delay $\tau$ (Fig.~\ref{fig:2ls}). The definition (\ref{defCorrF}) guarantees that time-tags corresponding to statistically independent outcomes between both SPDs give rise to $\alpha(w,\tau)=\beta(w,\tau)=\gamma(w,\tau)=1$ irrespective of $w$ and $\tau$. This is only the case when all detected photonic modes are occupied by pure Poisson noise. On the contrary, we reach sub-Poissonian statistics \cite{Mandel1979} when $\alpha(w,\tau)<1$, $\beta(w,\tau)<1$ and $\gamma(w,\tau)>1$. The sub-Poissonian character remains for all time bins with a size $w<w_0$, where $w_0$ is a particular time bin size, which implies that all the correlation functions in Eq.~(\ref{defCorrF}) become unity. Apart from this special choice of $w$, these correlation functions differentiate themselves. Each of them uniquely addresses different properties of the photonic state.


In the following we illustrate these properties with a simplified model allowing for continuous pumping of an ensemble of $N$ independent single-photon emitters and Poisson background noise that contributes to the optical field. To facilitate demonstration of the correlation function properties, we assume that the single-photon emitters contribute equally to the detected signal. However, all conclusions remain the same, even if emitters are of different brightness or couple differently to the detection system. Let $\eta$ and $\bar{n}$ denote the rate of emission from each single emitter and mean photon flux of the Poisson background noise, respectively. The correlation function $\alpha(w,\tau)$ provides useful information about the state in the limit of asymptotically decreasing $w$. Then, for $\tau=0$, we approach $\alpha(w,0) \approx 1-N/(N+\bar{n}/\eta)^2$, which depends on the flux of Poisson background noise with $\bar{n}$, while losses reducing both $\bar{n}$ and $\eta$ by the same factor do not impact $\alpha(w,0)$. In contrast, the correlation functions $\beta(w,0)$ and $\gamma(w,0)$ trivially approach unity with decreasing $w$ for any continuous wave photonic state, since in this limit the number of no-click time bins diverges.
Therefore, $\beta(w,0)$ and $\gamma(w,0)$ must be evaluated beyond this limit.
Assuming $0<\eta w\ll 1$, we obtain $\beta(w,\tau) \approx 1-(N \eta w)^2/4$ indicating that $\beta(w,\tau)$ is insensitive to Poisson noise, but is affected by losses and depends on the bin size $w$. Finally, in the limit $0<\eta w \ll 1$, we also obtain $\gamma(w,0)\approx 1+N \eta^2 w /\left[2(N\eta+\bar{n})\right]$, and hence a source with negligible noise $\bar{n} \ll \eta$ yields $\gamma(w,0) \approx 1+\eta w/2$, which scales linearly with $\eta w$ and is independent of the number of emitters. This example of a simple photonic state thus illustrates the diverse properties of $\alpha(w,\tau)$, $\beta(w,\tau)$ and $\gamma(w,\tau)$, that together advance the evaluation of the commonly used HBT detection scheme.

To evaluate physically relevant properties of single-photon emitters, we need to perform an analysis 
that is independent of the data processing. First, we define success $\bar{\gamma}(\tau)$ 
as the supremum
\begin{equation}
\bar{\gamma}(\tau) \equiv \sup_{w} \gamma(w,\tau).
\label{maxSuccess}
\end{equation}
Optimizing the success $\bar{\gamma}(\tau)$ enables applications in the development of single-photon emitters.
Second, we consider only the success probability $P_{10}(w,\tau)$ with a proper normalization and introduce the success rate $R_{w_0}(\tau)$ per calibrated time bin as 
\begin{equation}
\begin{aligned}
    R_{w_0}(\tau) \equiv \frac{P_{10}(w,\tau) w_0}{w},
    \label{Rw}
\end{aligned}
\end{equation}
where $w_0$ represents the size of the time bin implying a specific property in the measurement. Particularly, we allow for
$w_0$ that yields $\alpha(w_0,0)=\beta(w_0,0)=\gamma(w_0,0)=1$. To show that the normalization in Eq.~(\ref{Rw}) is not unique, we discuss a different choice of $w_0$ in Appendix A (section \ref{SM:correlationsF}). Because $P_{10}(w,\tau)$ is linear proportional to $w$ in the limit of weak photonic states with negligible two and higher photon detection probabilities, the definition in Eq.~(\ref{Rw}) ensures that $R_{w_0}(\tau)$ stays dimensionless and independent of $w$ in the limit of weak photonic states (Fig.~\ref{fig:figR}). Moreover, losses and Poisson background noise do not influence the calibration time bin $w_0$. Thus, these imperfections impact only $P_{10}(w,\tau)$ in Eq.~(\ref{Rw}) and $R_{w_0}(\tau)$ becomes as sensitive to them as the success probability $P_{10}(w,\tau)$.
Finally, we can introduce the smallest multiphoton error $\bar{\alpha}(\tau)$ and the smallest failure of emission $\bar{\beta}(\tau)$ as:
\begin{equation}
\begin{aligned}
  \bar{\alpha}(\tau)&\equiv \inf_{w} \alpha(w,\tau),\\
  \bar{\beta}(\tau)& \equiv \inf_{w} \beta(w,\tau),
\end{aligned}
\label{optABG}
\end{equation}
which correspond to properties that impose different requirements on engineering single-photon emitters as compared to $\bar{\gamma}(\tau)$ in Eq.~(\ref{maxSuccess}).
As an example, we note that $w_{max}$ such that $\gamma(w_{max}, \tau)=\bar{\gamma}(\tau)$ does not yield the minimum error $\bar{\alpha}(\tau)$ or minimum failure $\bar{\beta}(\tau)$. 
Thus, the threefold evaluation represents a trade-off between the different normalized quantities (\ref{maxSuccess}) and (\ref{optABG}). These correlation functions together with the success rate $R_{w_0}(\tau)$ provide an evaluation suitable for certification of single-photon emitters radiating a photonic state that is significantly affected by noise and losses~(Fig.~\ref{fig:figR}).


\section{Description of the experiment}

\begin{figure*}[htb]
\centerline {\includegraphics[width=\linewidth]{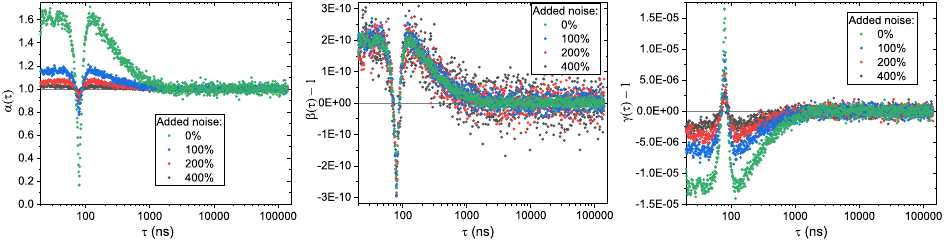}}
\caption{Plots of the correlation functions $\alpha(w,\tau)$, $\beta(w,\tau)$, and $\gamma(w,\tau)$ evaluated from the experimental data with a time-bin size of $w=1$~ns and various levels of added noise. Here the time axis is shifted by 80~ns for better illustration of the plotted functions. The horizontal line where the function value is equal to unity shows the classicality threshold: Values of the correlation functions $\alpha$, $\beta$ ($\gamma$) below (above) this line correspond to non-classical light. Note that $\alpha(\tau)$ for the selected time-bin size is identical to the second-order correlation function $g^{(2)}(\tau)$. Note also that the plots for $\beta(\tau)$ at different noise levels overlap, demonstrating invariance of $\beta(w,\tau)$ over added Poisson noise.}
\label{fig:abg_tau_noise}
\end{figure*}

\begin{figure*}[bht]
\centerline {\includegraphics[width=0.98\linewidth]{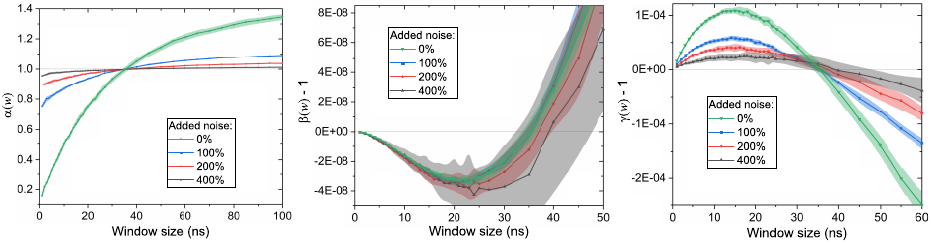}}
\caption{Plots of the correlation functions $\alpha(w,0)$, $\beta(w,0)$, and $\gamma(w,0)$ evaluated from the experimental data at various levels of added noise and used to extract $\bar{\alpha}$, $\bar{\beta}$, and $\bar{\gamma}$. Experimental data points are connected for clarity, and shaded regions show the error range. The horizontal line at $\alpha(w)=\beta(w)=\gamma(w)=1$ shows the classicality threshold. Note that the plots for $\beta(w)$ at different noise levels overlap, demonstrating invariance of $\beta(w,\tau)$ over added Poisson noise.}
\label{fig:abg_w_noise}
\end{figure*}

We experimentally determine the correlation functions $\alpha$, $\beta$, and $\gamma$ by evaluating correlations of the detection events from an HBT setup with incident emission from  single nitrogen-vacancy color centers (NVs) in diamond, established as well-understood and stable quantum emitters~\cite{Doherty2013}. In our electronic-grade bulk diamond sample, NV centers were created by nitrogen-ion implantation (energy 7.1~keV) with subsequent annealing at $800^\circ$C \cite{deLeon_prx19}. The implantation dose was $10^8$ ions/cm$^2$ that yields the formation of isolated NVs with a density $\approx 10^7$/cm$^2$. We optically excite a single NV with a 532~nm continuous-wave (CW) laser (0.8~mW power before the 100x objective with ${\rm NA}=0.95$) and collect fluorescence in a confocal configuration. The level of Poissonian background noise we control independently by adding light from a CW 637~nm laser to the detection path through a beam-splitter located right before the beam-splitter of the HBT configuration. Afterwards, the total optical field containing fluorescence from the NV and a controllable fraction of background noise is coupled to a single-mode optical fiber, split with a fiber beam-splitter and detected with two SPDs (Excelitas SPCM-AQRH-13), thus forming the HBT setup. Each detection event is recorded with 4~ps resolution using a PicoHarp 300 from PicoQuant operating in the time-tagged mode. This allows for calculating the three correlation functions $\alpha(w,\tau)$, $\beta(w,\tau)$, and $\gamma(w,\tau)$ in post-processing of the recorded data. For our analysis, each data set was recorded during one hour to ensure enough events with high-quality statistics and reliable estimation of the measurement error. The signal from one of the SPDs is artificially delayed such that the zero time shift between channels corresponds to $\tau=80$~ns.

We collect photon-counts from an isolated NV centre in several independent measurement rounds. In the first round, the NV center is excited by the pump 532-nm laser, and we collect fluorescence signal without artificial background. In the next three measurement rounds, we add various amounts of background noise by adjusting the power of the 637~nm laser to 100, 200, and $400\%$ of the pure fluorescence signal, respectively.

We process the experimental data by dividing the independent measurement rounds into time bins of size $w$ and applying Eqs.~(\ref{defCorrF}) to calculate correlations of click and/or no-click events. Note that according to the definition, a click event is the presence of {\it at least} one click in a time bin, while a no-click event is the absence of clicks in a time bin. Therefore, we do not distinguish events with just one or multiple clicks in a single time bin. Such definition explicitly allows measuring the correlation functions (\ref{defCorrF}) by non-photon-number-resolving detectors. In contrast, the (Glauber's) second-order correlation function $g^{(2)}(\tau)$ can only be measured, when the interval between photon arrivals is substantially larger than the dead time of the detection system.

The corresponding plots of the correlation functions $\alpha(w,\tau)$, $\beta(w,\tau)$, and $\gamma(w,\tau)$ for a fixed time bin $w=1$~ns are shown in Fig.~\ref{fig:abg_tau_noise} for the four cases of varying Poisson noise. According to discussions in section~\ref{sec_definitions}, the plots of $\beta(\tau)$ for different cases coincide (the signal-to-noise ratio varies though), demonstrating that $\beta(w,\tau)$ is insensitive to Poisson noise. Unsurprisingly, the plots for $\alpha(w,\tau)$ resemble those of the familiar second-order correlation function $g^{(2)}(\tau)$. The highest photon count rate in the experiment does not exceed 100~kCount/s at each detector, whereas the dead time is dominated by that of the PicoHarp 300 counting device ($<95$~ns). In this regime and given $w=1$~ns, the probability of having several photons within a single time bin is negligible and $\alpha(w,\tau)$ is nearly identical to $g^{(2)}(\tau)$. A more detailed discussion of the relation between correlation functions~(\ref{defCorrF}) and the Glauber quantum correlation functions is given in Appendix A (section~\ref{sec_probabilities}).

Similarly, in Fig.~\ref{fig:abg_w_noise} we plot the correlation functions $\alpha(w,\tau)$, $\beta(w,\tau)$, and $\gamma(w,\tau)$ for the four cases of varying Poisson noise and, contrary to Fig.~\ref{fig:abg_tau_noise}, for the fixed zero time shift $\tau$ between channels. These plots show the evolution of the correlation functions over a span of time bins $w$, and allow extracting the extreme values of error, failure, and success parameters $\bar{\alpha}$, $\bar{\beta}$, and $\bar{\gamma}$, respectively, according to Eqs.~(\ref{maxSuccess}) and (\ref{optABG}). The measured correlation functions for the relevant cases of $N$ independent quantum emitters and signal from a single emitter recorded with varying loss are summarized in Appendix D section~\ref{sec_extraplots} (Fig.~\ref{fig:abg_tau_merge}).

\begin{figure}[t!]
\centerline {\includegraphics[width=0.99\linewidth]{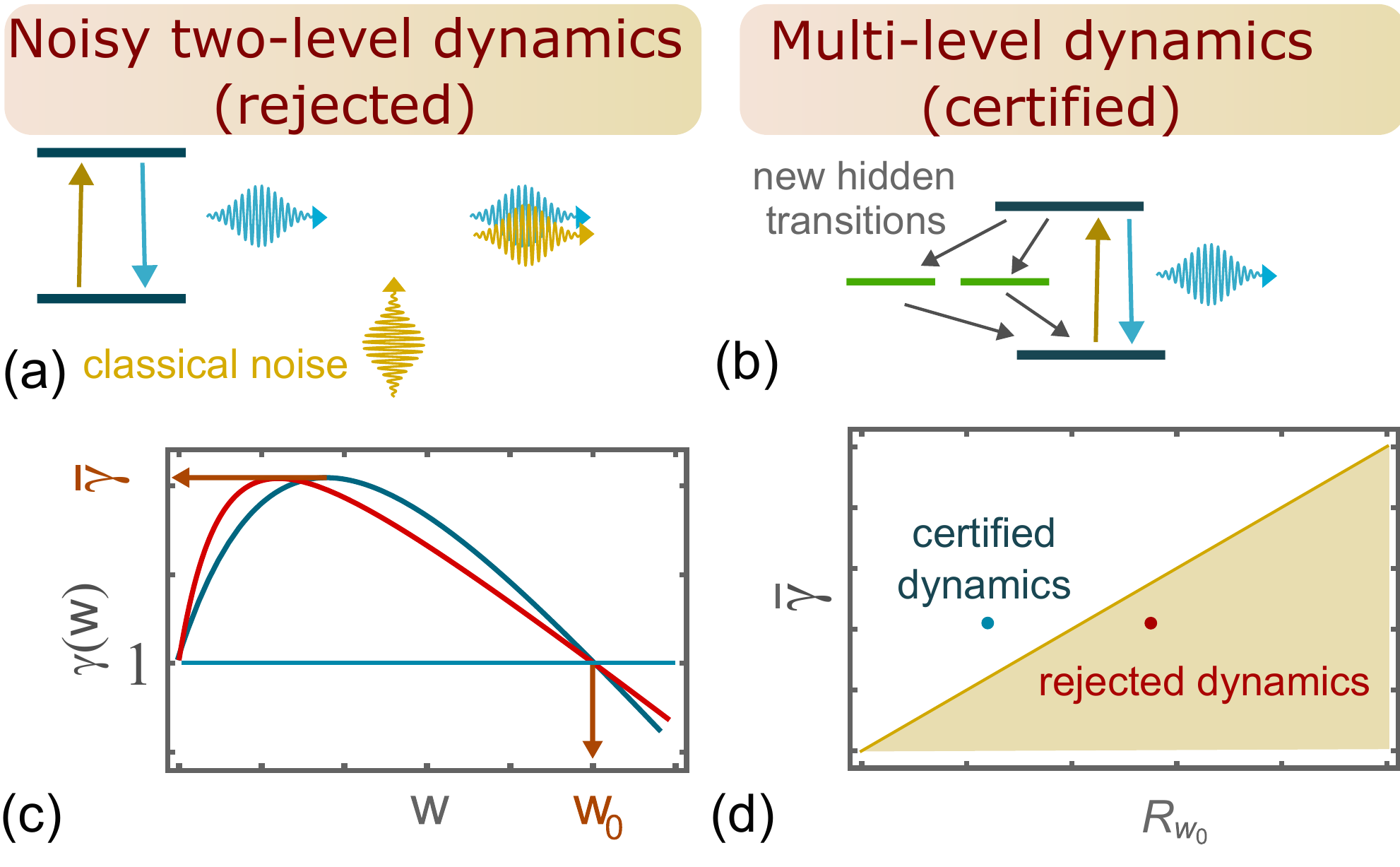}}
\caption{(a) Illustration of a model consisting of an emitter with dynamics between two energy levels and a noisy background. The transition from an excited state to a ground state radiates a photon (the blue wave) that is mixed with classical background noise (the yellow wave).  (b) Illustration of an emitter with multi-level dynamics. A direct transition from an excited state to a ground state radiates a photon (the blue wave). The indirect transition through other states (green lines) happens without emission of light. (c) The correlation function $\gamma(w)$ for various time bin sizes $w$ evaluated for an example emitter with multi-level dynamics (blue line) and a two-level emitter with noisy background (red line). One can not distinguish these two situations relying only on the maximum $\bar{\gamma}\equiv \max_{w}\gamma(w)$ and $w_0$ such that $\gamma(w_0)=1$.  (d)~Certification of the multi-level dynamics employing $\bar{\gamma}\equiv \max_w \gamma(w)$ and the parameter $R_{w_0}$ for rejection of any two-level dynamics with noise. All photonic states emerging in the noisy two-level dynamics reach only the plotted yellow region, covered by the orange threshold. The blue (red) point corresponds to the example of multi-level dynamics (noisy two-level dynamics) yielding $\gamma(w)$ depicted in (c) by the blue (red) line. The blue point surpasses the threshold, and therefore we certify the multi-level dynamics in this case.}
  \label{fig:figRejection}
\end{figure}

\section{Certification of emitters}

Quantum emitters of light posses a discrete energy level structure and the dynamics between these levels are reflected in the statistics of the emitted light. In this section, we apply the threefold evaluation introduced in section \ref{sec_definitions} to our experimental data obtained with a single NV centre. With this approach, we prove that the emitters have to be described with a multilevel structure rather than a simple two-level system mixed with classical noise (Fig.~\ref{fig:figRejection}).

We note a distinct difference to the commonly used approach based on the evaluation of Glauber's second-order correlation function 
\begin{equation}
    g^{(2)}(\tau)= \lim_{w\rightarrow 0}\alpha(w,\tau).
\end{equation}
The presence of both bunching and anti-bunching in $g^{(2)}(\tau)$ does not guarantee immediately that the emitter has more than two energy levels. Except for noise with a Poisson distribution, any classical background noise mixed with the signal from a two-level system results in a second-order correlation function possessing both bunching and anti-bunching. In order to prove emission from a multi-level system, the measured amount of bunching and anti-bunching has to be of a certain minimum level (confer Eq.~(\ref{SM:critA}) in Appendix C). In contrast to this, in this section we aim at application of a criterion based on the correlation function $\gamma(w)$, Eq.~(\ref{SM:critG}). With this, we certify a nontrivial energy level structure of an emitter based on a property that is interpreted as a success.

In order to capture the emitter dynamics in our model, we expand the correlation functions $\gamma$ in Eq.~(\ref{defCorrF}) into a Taylor series of the normally ordered moments of $a^{\dagger}$ and $a$ (see Appendix A for more details) and then use the regression theorem \cite{Loudon2000} to link those moments with the emitter dynamics. Here, the model system consists of on optical ground and excited state with allowed radiative relaxation and off-resonant (incoherent) pumping. Assuming $\tau=0$ for simplicity, the emitted photonic state exhibits trivial limits $\alpha(w,0) \ll 1$ and $\beta(w,0)\approx \gamma(w,0) \approx 1$ for small $w$. Increasing $w$ leads to higher $\alpha(w,0)$ due to multiple excitations of the two-level system within the time bin $w$. In this case, advantageously the failure $\beta(w,0)$ decreases while the success $\gamma(w,0)$ increases. For light emitted from an ideal two-level system, the correlation function $\gamma(w,0)$ obeys the  limit $\gamma(w,0)>1$ for $w>0$. Thus, a photonic state from a two-level system exhibits $w_0 \gg 1$ and, consequently, $R_{w_0}(0)\gg 1$. However, classical background noise mixed with the light emitted from an ideal two-level system causes that both $R_{w_0}(0)$ and $\bar{\gamma}(0)$ gain finite values. This is comparable with the measured data as depicted in Fig.~\ref{fig:thresW0}. We detail this analysis in Appendix C section \ref{SM:crit}.

\begin{figure}[t!]
\centerline {\includegraphics[width=0.99\linewidth]{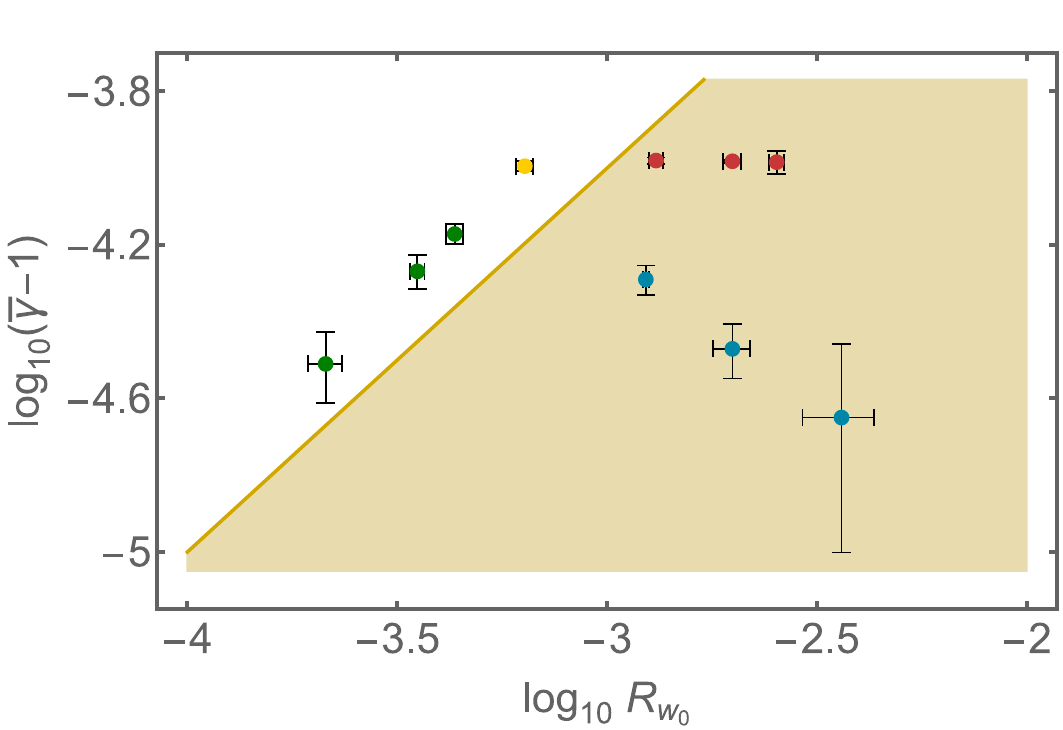}}
\caption{Certification of an emitter with a shelving state based on rejection of an emitter with two energy levels that classical background noise deteriorates. This certification relies only on $\bar{\gamma}\equiv \max_w \gamma(w)$ and normalized success rates $R_{w_0}$. The yellow region corresponds to such a pair of $\bar{\gamma}$ and  $R_{w_0}$ that can be achieved by a two-level system deteriorated by arbitrary classical background noise affecting an emitter with two levels of energy. On the contrary, all states above the yellow region can not be explained by this model. The yellow point represents this evaluation applied to a realistic photonic state radiated from the NV center in diamond. The sequence of blue points demonstrates a path that the photonic state follows in this diagram when added Poisson background noise increases the click rates by $100\%$, $200\%$ and $400\%$. The row of red points shows how the state moves in this diagram when we merge the measured time-tag data in two or more time windows to simulate emission from two, three and four independent NV centers. The sequence of green points depicts cases when additional losses $20\%$, $37\%$ and $68\%$ affect the measured photonic state.}
\label{fig:thresW0}
\end{figure}

We next apply this methodology to certify a nontrivial multi-level structure of the NV centre, known to have a (third) shelving state enabling a non-radiative relaxation path~\cite{Doherty2013}, and therefore is a good system to demonstrate the model rejection procedure. We use the data obtained with the NV center (see previous section and Appendix section~\ref{sec_extraplots}) and allow for realistic experimental conditions when (i) background Poisson noise deteriorates the radiated photonic state, when (ii) the emission occurs from more than one emitter, or when (iii) the photonic state is affected by losses. We test our emitter for being a two-level system with classical background noise. In Fig.~\ref{fig:thresW0}, we plot the corresponding rejection criterion based on the parameters $\bar{\gamma}(0)$ and $R_{w_0}(0)$ with $w_0$ implying $\gamma(w_0)=1$ as derived in Eq.~(\ref{SM:critG}).
The yellow shaded area corresponds to emitted photonic states that can be generated from systems with two energy levels and mixed with classical background noise. From each set of our experimental data we extract the values $\bar{\gamma}(0)$ and $R_{w_0}(0)$ and plot these in the same Fig.~\ref{fig:thresW0} for the cases of (i) added Poissonian noise (blue data points), (ii) $N$ independent and equally bright emitters (red data points) and (iii) loss affecting the measured photonic state (green data points). The data points in the white region fall into the model-rejection region, for which we can confidently reject any model consisting of two energy levels and classical background noise. The points in the yellow region are not model rejected and therefore, from the respective measurement we cannot tell with confidence whether the photonic state is radiated from a two-level system affected by classical noise or whether a multi-level energy structure is involved. Data points with added Poissonian noise (i, blue) and $N>1$ emitters (ii, red) fall in the yellow shaded region, while all cases of added loss (iii, green) are in the model-rejection region. Note that the ambiguity in the definition of $w_0$ in Eq.~(\ref{Rw}) principally enables derivation of more criteria that differ themselves by their requirement on the emitted state.

\section{Conclusion and outlook}

In this article, we suggest and experimentally implement an advanced evaluation methodology to certify non-trivial energy level dynamics of quantum light emitters. Our threefold evaluation strategy is based on the standard Hanburry-Brown and Twiss setup, and can therefore readily be applied for many physical systems. We define three correlation functions based on the possible combinations of clicks and no-clicks of the two detectors in the HBT setup and we illustrate their properties using a simple example with N independent emitters and Poissonian background noise. We apply the methodology to light emitted from NV centres in diamond and we investigate the the correlation functions with (i) additional background Poisson noise, (ii) when the emission occurs from more than one emitter, or (iii) when the photonic state is affected by losses. Finally, our analysis certifies that the dynamics of a single NV centre cannot be explained with only two energy levels and classical background noise.



As an outlook, we imagine application of our evaluation strategy to other quantum light emitters with unknown or more complex level dynamics, and where e.g. novel excitation schemes are required to control and improve the properties of emitted photonic states~\cite{Thomas2021, Knall2022, Wei2022}. Further investigations will aim at certification that the dynamics of an emitter are more complex than the ones of a three level system including a shelving state. First, we will analyse the  radiation of emitters with transitions between many energy levels \cite{Huber2017,Masters2023}. Second, we will explore in more detail the properties of emitted light radiated from several emitters and investigate the impact of light-matter dynamics between the emitters. For this case, we will consider an extended detection scheme based on a general splitting network comprising several beamsplitters and detectors~\cite{Straka2018,Lachman2019}, while alternatively a homodyne detection can be included for the analysis~\cite{Hacker2019,Magro2023}. Both detection schemes allow exploring larger Hilbert spaces, thus extracting more information about quantum light radiation processes compared to the HBT scheme. Such a scheme will have potential application also in many body systems emitting light\cite{Mahmoodian2020}, topological waveguide QED \cite{Bello2019} or optomechanics \cite{Iorsh2020,Sedov2020}.


\section*{Acknowledgements}
LL acknowledges the project 23-06015O and and RF was supported by the project 23-06308S, both of the Czech Science Foundation. RF also acknolwedges the funding from the MEYS of the Czech Republic (Grant Agreement 8C22001). Project SPARQL has received funding from the European Union’s Horizon 2020 Research and Innovation Programme under Grant Agreement no. 731473 and 101017733 (QuantERA). IPR, MB, ULA and AH acknowledge financial support from the Horizin 2020 twinning project NONGAUSS (grant agreement ID: 951737), the Novo Nordisk Foundation through the project Biomag (NNF21OC0066526), the Danish National Research Foundation (DNRF) through the center for Macroscopic Quantum States (bigQ, Grant No. DNRF0142), and the EMPIR program co-financed by the Participating States and from the European Union’s Horizon 2020 research and innovation programme via the QADeT project (Grant No. 20IND05).

\section{Appendix}

\subsection*{Appendix A: Simulating response of a detector in HBT configuration}\label{SM:correlationsF}
The HBT configuration employs two SPDs to measure light passing through a beam splitter (BS). A single SPD measures the probability $P_1(w)$ of at least one photon transmitted (reflected) in a time bin. Simultaneously, a SPD allows detection of the complementary probability $P_0(w)$ of no photons in a time bin. Considering both SPDs in HBT, allows us to define the probability $P_{11}(w,\tau)$ of detecting at least one transmitted photon and, simultaneously, at least one reflected photon in time bins having size $w$ and separated by time $\tau$. To complete description of detection in the HBT scheme, we further define the probability $P_{10}(w,\tau)$ of at least one transmitted photon and no reflected photons and, finally, the probability $P_{00}(w,\tau)$ of no photons impinging the detectors in the respective time bins. The introduced probabilities are not independent but they follow the relations
\begin{equation}
\begin{aligned}\label{SM:probRelations}
    P_0(w) &=1-P_1(w)\\
    P_{10}(w,\tau) &=P_1(w)-P_{11}(w,\tau)\\
    P_{00}(w,\tau) &=1-2P_1(w)+P_{11}(w,\tau).
\end{aligned}
\end{equation}
We can use these probabilities to define three correlation functions $\alpha(w,\tau)$, $\beta(w,\tau)$ and $\gamma(w,\tau)$ following as:
\begin{equation}\label{SM:corrFs}
\begin{aligned}
\alpha(w,\tau) &\equiv \frac{P_{11}(w,\tau)}{P_1(w)^2}\\
\beta(w,\tau) &\equiv \frac{P_{00}(w,\tau)}{P_1(w)P_0(w)}\\
\gamma(w,\tau) &\equiv \frac{P_{10}(w,\tau)}{P_1(w)P_0(w)}.
\end{aligned}
\end{equation}
The correlation function $\alpha(w,\tau)$, $\beta(w,\tau)$ and $\gamma(w,\tau)$ employ the probabilities $P_1(w)$ and $P_0(w)$ for normalization. Simultaneously, we allow for a different normalization of the probability $P_{10}(w)$ and introduce the parameter $R_{w_0}$ following as:
\begin{equation}\label{SM:Rpar}
R_{w_0}=\frac{P_{10}(w,\tau) w_0}{w},
\end{equation}
where $w_0$ corresponds to the size of a time bin implying a specific property in the measurement. Example choices of $w_0$ lead to $\gamma(w_0)=1$ or $\gamma(w_0)=\max_{w}\gamma(w)$.
The correlation functions in Eq.~(\ref{SM:corrFs}) and the parameter $R_{w_0}$ in Eq.~(\ref{SM:Rpar}) can be directly measured in the HBT detection scheme. However, theoretical simulation of the response in this detection is based on moments of the normally ordered moments of the creation operator $a^{\dagger}$ and the annihilation operator $a$ \cite{Blow1990}. Further, we summarize the theory of time dependent detection \cite{Blow1990,Fischer2016}. We first consider measurement of classical, distinguishable particles and then switch to the quantum description.

\subsubsection{Classical picture - beam of distinguishable particles}
We allow for a beam of $m$ classical particles with temporal distribution determined by the probability density function $\calT(t_1,...,t_m)$, where the argument $t_i$ represents time when $i$th particle arrive at a detector \cite{Kelley1964}. Further, we introduce the marginal distribution $\calM_{ij}(t_i,t_j)$ of $i$th and $j$th particle according to
\begin{equation}
    \calM_{ij}(t_i,t_j)\equiv \int \mathrm{d \mathbf{t}_{\overline{ij}}}\calT(t_1,...,t_m),
\end{equation}
where $\mathrm{d}\mathbf{t}_{\overline{ij}}$ symbolizes that the integration is taken over arrival times of all particles except $i$th and $j$th particle. The marginal distribution for $i$th particle $\calM_{i}(t_i)$ is defined as
\begin{equation}
    \calM_i(t_i)=\int \mathrm{d}t_j \calM_{ij}(t_i,t_j).
\end{equation}
In the following, we allow for a specific case of $\calT(t_1,...,t_m)$ that is symmetric in exchanging any two particles, and therefore the marginal distributions $\calM_{i}(t_i)$ and $\calM_{i,j}(t_i,t_j)$ hold
\begin{equation}
\begin{aligned}
\calM_{i}(t_i) &= \calM^{(1)}(t_i)\\
 \calM_{i,j}(t_i,t_j) &= \calM^{(2)}(t_i,t_j)
\end{aligned}
    \label{margPDF}
\end{equation}
identically for any $i$ and $j$.
Based on this conjecture, we can determine the mean number of particles $\langle n\rangle_w$ and particle pairs $\langle \binom{n}{2}\rangle$ in a time bin characterized by its size $w$ according to
\begin{equation}
\begin{aligned}
    \langle n\rangle_w &=\sum_i \int_{\tau}^{\tau+w}\mathrm{d}t_i \calM_i(t_i)\\
    &=m \int_{\tau}^{\tau+w}\mathrm{d}t \calM^{(1)}(t)\\
    \biggl< \binom{n}{2}\biggr>_w &=\frac{1}{2}\sum_{i\neq j}\int_{\tau}^{\tau+w}\mathrm{d}t_1\int_{\tau}^{\tau+w}\mathrm{d}t_2 \calM_{i,j}(t_i,t_j)\\
    &=\binom{m}{2}\int_{\tau}^{\tau+w}\mathrm{d}t_1\int_{\tau}^{\tau+w}\mathrm{d}t_2 \calM^{(2)}(t_1,t_2).
\end{aligned}
\end{equation}
Further, we rearrange the formula for the mean number of particle pairs so that it can be exploited later for indistinguishable particles. Let $P_{i,j}(w)$ denote a probability that $i$th and $j$th particle impinge in the same time bin, i.e.
\begin{equation}
\begin{aligned}
    P_{i,j}(w)=\int_{\tau}^{\tau+w}\mathrm{d}t_1\int_{\tau}^{\tau+w}\mathrm{d}t_2 \calM^{(2)}(t_i,t_j).
\end{aligned}
\end{equation}
Since these particles are distinguishable, we can determine when the $i$th particle arrives before $j$th particle. Let $P_{i\rightarrow j}(w)$ be the probability that this happens when these particles arrive at the detector in the same time bin. We thus obtain
\begin{equation}
    P_{i\rightarrow j}(w)=\int_{\tau}^{\tau+w}\mathrm{d}t_i\int_{t_i}^{\tau+w}\mathrm{d}t_j \calM^{(2)}(t_i,t_j).
\end{equation}
Due to the symmetry in particle exchange, $P_{j\rightarrow i}=P_{i\rightarrow j}$, and therefore the probability $P_{i,j}(w)$ can be expressed as
\begin{equation}
\begin{aligned}
  P_{i,j}(w)&=P_{i\rightarrow j}(w)+P_{j\rightarrow i}(w)\\
  &=2 \int_{\tau}^{\tau+w}\mathrm{d}t_1\int_{t_1}^{\tau+w}\mathrm{d}t_2 \calM^{(2)}(t_1,t_2).
\end{aligned}
\end{equation}
Putting all together, we can conclude that
\begin{equation}
\begin{aligned}
    \biggl< \binom{n}{2}\biggr>_w&=m(m-1) \\
    &\times \int_{\tau}^{\tau+w}\mathrm{d}t_1\int_{t_1}^{\tau+w}\mathrm{d}t_2 \calM^{(2)}(t_1,t_2),
\end{aligned}
\end{equation}
where the integration is performed over $t_1$, which stands for the arrival time of the earlier particle, and over $t_2$, which corresponds to the impinging time of the later particle. This time order becomes important when we deal with the quantum description of indistinguishable particles.


\subsubsection{Quantum picture - beam of indistinguishable photons}
The quantum description in discrete variables employs the Fock state basis to identify any photonic state. Let us consider a field composed of $M$ temporal modes. The theory of propagating photonic field introduces the annihilation operator $a(t)$ and the creation operator $a^{\dagger}(t)$ that obey the commutation relation $\left[a^{\dagger}(t_1),a(t_2)\right]=\delta(t_1-t_2)$. The physical description is completed by a set of orthonormal complex functions $\phi_i(t)$ determining temporal shapes of the modes \cite{Blow1990}. The Fock states are defined as a result of acting the creation operators on the vacuum. For example, the Fock state $|1\rangle$ is given by \cite{Raymer2020}
\begin{equation}
    \begin{aligned}
        |1\rangle & \equiv \int \mathrm{d}t f_1(t)a_i^{\dagger}(t) |0\rangle,
    \end{aligned}
\end{equation}
where the function $f_1(t)\equiv \sum_{i=1}^M\tau_i \phi_i(t)$ with complex amplitudes $\tau_i$ fulfills $\int |f_1(t)|^2\mathrm{d}t=1$. 
Analogously, a general Fock state $|m\rangle$ in this description gains
\begin{equation}
    |m\rangle \equiv \frac{1}{\sqrt{m!}}\int \mathrm{d}t_1...\mathrm{d}t_m f_m(t_1,...,t_m)a^{\dagger}(t_1)...a^{\dagger}(t_m)|0\rangle,
    \label{def:FockN}
\end{equation}
where $f_m(t_1,...,t_m)$  is invariant under any permutation of its arguments and obey $\int \mathrm{d}t_1...\mathrm{d}t_m |f_n(t_1,...,t_m)|^2=1$. Further, we define the functions $\calM^{(1)}(t)$ and $\calM^{(2)}(t_1,t_2)$ as
\begin{equation}
\begin{aligned}
    \calM^{(2)}(t_1,t_2)&\equiv \int \mathrm{d}t_3...\mathrm{d}t_n |f_m(t_1,...,t_m)|^2\\
    \calM^{(1)}(t)&\equiv \int \mathrm{d}\tau \calM^{(2)}(t,\tau).
\end{aligned}
\end{equation}
An intensity detector in Hanbury-Brown and Twiss configuration operating in the wideband limit with very high temporal resolution \cite{Yurke1985} measures the normally ordered moments \cite{Glauber1963a} following as:
\begin{equation}
\begin{aligned}
    \langle m|a^{\dagger}(t_1)a^{\dagger}(t_2)a(t_2)a(t_1)|m\rangle &=m(m-1) \calM^{(2)}(t_1,t_2)\\
    \langle m| a^{\dagger}(t)a(t)|m\rangle &=m \calM^{(1)}(t),
\end{aligned}
\end{equation}
where we allow for detection of the Fock state $|m\rangle$ defined in (\ref{def:FockN}).
Because $\calM^{(1)}(t)$ and $\calM^{(2)}(t_1,t_2)$ correspond to a probability density function of observing a photon in time $t$ and observing photons in time $t_1$ and $t_2$, respectively, we can exploit the analogy with distinguishable particles and apply the same procedure to integrate over a time bin of size $w$. Thus, the Fock state $|m\rangle$ induces the following mean number of photons in a time bin: 
\begin{equation}
    \langle n\rangle_w= m \int_{\tau}^{\tau+w}\mathrm{d}t \calM^{(1)}(t)
    \label{FirstMoment}
\end{equation}
and the following mean number of photon pairs:
\begin{equation}
\begin{aligned}
    \biggl< \binom{n}{2}\biggr>_w  &=m(m-1) \\
    &\times \int_{\tau}^{\tau+w}\mathrm{d}t_1\int_{t_1}^{\tau+w}\mathrm{d}t_2 \calM^{(2)}(t_1,t_2).
\end{aligned}
\label{secondMoment}
\end{equation}
To determine the moments $\langle n\rangle_w$ and $\langle \binom{n}{2}\rangle_w$ for a general photonic state, we employ the representation in the Fock state basis and apply (\ref{FirstMoment}) and (\ref{secondMoment}) to the Fock states in the basis. Consequently, the moments $\langle n\rangle_w$ and $\langle \binom{n}{2}\rangle_w$ works out to be
\begin{equation}
\begin{aligned}
    \langle n\rangle_w &= \int_{\tau}^{\tau+w}\mathrm{d}t \langle n(t)\rangle\\
    \biggl< \binom{n}{2}\biggr>_w   &=\int_{\tau}^{\tau+w}\mathrm{d}t_1\int_{t_1}^{\tau+w}\mathrm{d}t_2 G^{(2)}(t_1,t_2),
\end{aligned}
\label{finalExp}
\end{equation}
where $\langle n(t)\rangle=\langle a^{\dagger}(t)a(t)\rangle$ and $G^{(2)}(t_1,t_2)=\langle a^{\dagger}(t_1)a^{\dagger}(t_2)a(t_2)a(t_1)\rangle$ is the Glauber's second-order correlation function \cite{Glauber1963,Sudarshan1963}. 
Let us note that the time $t_1$ and $t_2$ in integration (\ref{finalExp}) keep the normal temporal ordering, which is required in definition of $G^{(2)}(t_1,t_2)$, and therefore the identities in (\ref{finalExp}) are consistent with the theory of photodetection \cite{Loudon2000}. We can extend this for the mean number of tuples with $m$ photons according to
\begin{equation}
\begin{aligned}
    \biggl< \binom{n}{m}\biggr>_w   &=\int_{\tau}^{\tau+w}\mathrm{d}t_1 \int_{t_1}^{\tau+w}\mathrm{d}t_2...\int_{t_{m-1}}^{\tau+w}\mathrm{d}t_m \times\\
    & \times G^{(m)}(t_1,...,t_m),
\end{aligned}
\label{SM:mG}
\end{equation}
where we integrate over the ordered tuple of time $(t_1,...,t_m)$ such that $t_{k-1}<t_k$ for all $k \in \left\{2,...,m\right\}$. We formally simplify (\ref{SM:mG}) by introducing the operator $A(w)\equiv \int_{\tau}^{w+\tau} \mathrm{d}t\ a(t)$ and its Hermite conjugate operator $A^{\dagger}(w)\equiv \int_{\tau}^{w+\tau} \mathrm{d}t\ a^{\dagger}(t)$, which allows us to express
\begin{equation}
\biggl< \binom{n}{m}\biggr>_w=\frac{1}{m!}\langle \left[A^{\dagger}(w)\right]^m A(w)^m\rangle.
\label{SM:momentsA}
\end{equation}
The identity (\ref{SM:momentsA}) predicts a response of a photon-number resolving detector on a propagating photonic field \cite{Blow1990}.

\subsubsection{Click and no-click probabilities in HBT}
\label{sec_probabilities}
A SPD in HBT yields a click if at least one photon arrives. Detection in HBT configuration consists of two SPDs which allows measuring a probability $P_1$ of a click in a chosen SPD and a probability $P_{11}$ of coincidence click of both the SPDs. These probabilities work out to be \cite{Sperling2012}
\begin{equation}\label{SM:probExactly}
\begin{aligned}
 P_1(w) &= \langle \normord{1-e^{-\frac{T}{2} A^{\dagger}(w)A(w)}} \rangle \\
 P_{11}(w) &= \langle \normord{ \left(1-e^{-\frac{T}{2} A^{\dagger}(w)A(w)} \right)^2} \rangle,
\end{aligned}
\end{equation}
where the parameter $T$ simulates light collection and detection efficiency. Since $T$ is typically very small, click probabilities in (\ref{SM:probExactly}) can be approximated by leading terms in their Taylor series with respect to $T$, i.e.
\begin{equation}\label{SM:approxClick}
    \begin{aligned}
     P_1(w) &\approx \frac{T}{2} \langle A^{\dagger}(w) A(w)\rangle\\
     P_{11}(w) &\approx\left(\frac{T}{2}\right)^2 \langle \normord{\left[A^{\dagger}(w)A(w)\right]^2}\rangle.
    \end{aligned}
\end{equation}
Establishing $P_1(w)$ and $P_{11}(w)$ allows us to gain all the probabilities of the click and no-click events occurring in HBT due to the identities (\ref{SM:probRelations}).
Assuming the approximation (\ref{SM:approxClick}), we can express the correlation functions in Eq.~(\ref{SM:corrFs}) following as:
\begin{equation}
\begin{aligned}
\alpha(w) &\approx \widetilde{g}^{(2)}(w)\\
\beta(w) &\approx 1-\left[T \bar{n}(w)\right]^2\left[1-\widetilde{g}^{(2)}(w)\right]\\
\gamma(w) &\approx 1+T \bar{n}(w)\left[1-\widetilde{g}^{(2)}(w)\right],
\end{aligned}\label{SM:defCorrFs}
 \end{equation}
 where we employ
 \begin{equation}
 \begin{aligned}\label{SM:moments}
    \widetilde{g}^{(n)}(w)&\equiv \frac{\langle \normord{\left[A^{\dagger}(w)A(w) \right]^n}\rangle}{\left[\langle A^{\dagger}(w)A(w)\rangle \right]^n}\\
    \widetilde{n}(w) &\equiv \langle A^{\dagger}(w)A(w)\rangle.
 \end{aligned}
 \end{equation}
We can increase the accuracy of expressions for the correlation functions by considering more members in the Taylor series of (\ref{SM:probExactly}) with respect to small $T$. Particularly, we obtain
 \begin{equation}\label{SM:corFABG}
\begin{aligned}
\alpha(w) &\approx \widetilde{g}^{(2)}(w)+T \bar{n}(w)\left[\widetilde{g}^{(2)}(w)-\widetilde{g}^{(1)}(w)\widetilde{g}^{(3)}(w)\right]\\
\beta(w) &\approx 1-\left[T \bar{n}(w)\right]^2\left[1-\widetilde{g}^{(2)}(w)\right]\\
&+\left[T\bar{n}(w)\right]^3\left[3 \widetilde{g}^{(2)}(w)-\widetilde{g}^{(3)}(w)-2\right]\\
\gamma(w) &\approx 1+T \bar{n}(w)\left[1-\widetilde{g}^{(2)}(w)\right]\\
+\left[T \bar{n}(w)\right]^2 &\left\{2- 3 \widetilde{g}^{(2)}(w)-\left[\widetilde{g}^{(2)}(w)\right]^2+2\widetilde{g}^{(3)}(w)\right\},
\end{aligned}
 \end{equation}
which suggests that the accuracy of the approximations scales with $T\bar{n}(w)$. Employing more terms in the Taylor series of (\ref{SM:probExactly}) allows us to further increase the accuracy.


\subsection*{Appendix B: Detector response on model states}
Here, we apply the relations in (\ref{SM:probExactly}) to simulate a detector response on models of stationary photonic states with moments $\langle \normord{\left[a^{\dagger}(t)a(t)\right]^k}\rangle$ independent of time $t$ for any $k$. This regime is relevant for photonic states that emerge from continuous pumping a single-photon emitter.

\subsubsection{Classical noise}
Any classical noise can be represented as a stochastic mixture of the coherent states \cite{Glauber1963,Sudarshan1963}. When only the phase undergoes random changes but the intensity of the noise remains stable, the noise approaches the Poisson limit, which gives rise to the probabilities $P_1(w)=1-\exp(-T\bar{n}w/2)$ and $P_{11}(w)=\left[1-\exp(-T\bar{n}w/2)\right]^2$, where $T\bar{n}w$ quantifies the mean number of photons in a time bin. A general classical noise produces the click probabilities that depend also on the temporal correlation of the noise. In the regime $w T \ll 1$, only the moments $\langle A^{\dagger}(w)A(w)\rangle$ and $\langle \normord{\left(A^{\dagger}(w)A(w)\right)^2}\rangle$ affect significantly the probabilities $P_1(w)$ and $P_{11}(w)$, and therefore we identify any classical noise in this limit only by two moments
\begin{equation}
    \begin{aligned}
     \bar{n} &\equiv \langle a^{\dagger}(t)a(t)\rangle\\
     G^{(2)} \left(\tau \right)&\equiv \langle a^{\dagger}(t)a^{\dagger}(t+\tau)a(t+\tau)a(t)\rangle.
    \end{aligned}
    \label{SM:defCl}
\end{equation}
This implies the click probabilities formally achieve
\begin{equation}
    \begin{aligned}
        P_1(w) &\approx \frac{\eta w}{2}\bar{n}\\
        P_{11}(w)& \approx \left(\frac{\eta}{2}\right)^2 \widetilde{g}(w)\bar{n}^2,
    \end{aligned}
    \label{SM:classInEq}
\end{equation}
where we employed 
\begin{equation}\label{SM:noiseG2w}
    \widetilde{g}(w)\equiv \frac{1}{\bar{n}^2} \int_{0}^{w}\mathrm{d}t_1 \int_{t_1}^{w}\mathrm{d}t_2 G^{(2)}(|t_2-t_1|).
\end{equation}
Because classical noise exhibits $G^{(2)}(\tau)$ that drops for $\tau>0$, we can establish
\begin{equation}
    \widetilde{g}(w) \in \left\langle w^2, g w^2\right)
    \label{SM:defgw}
\end{equation}
with $g\equiv G^{(2)}(0)/\bar{n}^2$. The lower limit in Eq.~(\ref{SM:defgw}) is achieved by the Poisson noise and the upper limit gets saturated when $G^{(2)}(\tau)$ approaches the constant function in $\tau \leq w$.

\subsubsection{Above-band pumping a two-level system} \label{Sm:off2LS}
Let us consider a two-level system with an excited state $|e\rangle$ and a ground state $|g\rangle$. Above-band pumping the two-level system causes incoherent dynamics of the density matrix $\rho=\rho_e(t) |e\rangle \langle e|+\rho_g(t) |g\rangle \langle g|$. The evolution is described by the rate equations:
\begin{equation}
    \begin{aligned}
        \frac{d}{dt}\rho_g(t) &=-\kappa_p \rho_g(t)+\kappa_r \rho_e(t)\\
        \frac{d}{dt}\rho_e(t) &=\kappa_p \rho_g(t)-\kappa_r \rho_e(t),
    \end{aligned}
\end{equation}
where $\kappa_p$ represents rates of excitation and $\kappa_r$ corresponds to rates of spontaneous decay. Let us assume that the evolution of two-level system starts from the ground state. Then, the probability $\rho_e(t)$ of exciting the two-level system at time $t$ works out to be
\begin{equation}\label{SM:rhoTLS}
    \rho_e(t)=\frac{\kappa_p}{K}\left[1-e^{-K t}\right],
\end{equation}
where $K=\kappa_p+\kappa_r$. The regression theorem \cite{Loudon2000} yields $\langle n\rangle=\kappa_r \kappa_p/K$ and $G^{(2)}(\tau)=\left[\kappa_r \kappa_p/K\right]^2\left[1-\exp(-K t)\right]$ for this dynamics. Thus, we obtain
\begin{equation}
    \begin{aligned}
    P_1(w) & \approx \frac{T \kappa_r \kappa_p w}{2 K}\\
    P_{11}(w) & \approx \frac{T^2 \kappa_p^2 \kappa_r^2 \left[2(1-K w)-2e^{-K w}+K^2 w^2\right]}{4K^3}.
    \end{aligned}
    \label{SM:prob2LS}
\end{equation}
We can check numerically that the probabilities in Eq.~(\ref{SM:prob2LS}) imply the correlation function $\alpha(w)$ defined in Eq.~(\ref{SM:corrFs}) obeys $\alpha(w)<1$ for any $w$.

Mixing the light from a two-level system with Poisson background noise increases the error $\alpha(w)$ without gaining $\alpha(w)>1$ in the regime of reliable $w$ and reduces the success $\gamma(w)$. On the contrary, the failure $\beta(w)$ remains without any change. Mixing the light with general classical background noise, defined in the limit $T w\ll 1$ by click probabilities in Eq.~(\ref{SM:classInEq}), gives rise to approximate correlation functions following as:
\begin{equation}
    \begin{aligned}
    \alpha(w) &\approx 1+\frac{f_2(w)-1+\left[\widetilde{g}(w)-1\right](\bar{n}/\eta)^2}{(1+\bar{n}/\eta)^2}\\
    \beta(w) &\approx 1-\frac{\eta^2 w^2 T^2}{4}\\
    &\times \left[1-f_2(w)+(\widetilde{g}(w)-1)(\bar{n}/\eta)^2\right]\\
    \gamma(w) &\approx 1+\frac{\eta w T}{2(1+\bar{n}/\eta)}\\
    &\times \left[1-f_2(w)+(1-\widetilde{g}(w))(\bar{n}/\eta)^2\right].
     \label{SM:corr2LSnoise}
    \end{aligned}
\end{equation}
where $\bar{n}$ and $\widetilde{g}(w)$ characterize the noise contributions, $\eta\equiv \kappa_r \kappa_p/(\kappa_p+\kappa_r)$ and $f_2(w)$ reads as
\begin{equation}
f_2(w)\equiv \frac{\left[2(1-K w)-2e^{-K w}+K^2 w^2\right]}{4K w^2}
\end{equation}
with $K=\kappa_p+\kappa_r$.
The correlation functions in Eq.~(\ref{SM:corr2LSnoise}) become one unitedly for a given time bin with size $w_0$ if
\begin{equation}
\bar{n}/\eta=\frac{\sqrt{2\left[\exp(-K w_0)-1+K w_0\right]}}{K w_0 \sqrt{\widetilde{g}(w_0)-1}}.
\label{SM:w0Rel}
\end{equation}
Apart from the case of the Poisson background noise with $\widetilde{g}(w)=1$, we can deduce $w_0$ from (\ref{SM:w0Rel}) and establish the unnormilized success $R$ defined as $R\equiv P_{10}(w) w_0/w$ to complete the evaluation of these model states.

\subsubsection{Above-band pumping a three-level system with shelving state} \label{Sm:off3LS}
We extend the analysis of the two-level system by considering a dark transition from the excited state $|e\rangle$ to the ground state $|g\rangle$ through a shelving state $|s\rangle$. In a case of the off-resonant pumping, the evolution of the density matrix $\rho=\rho_e(t)|e\rangle\langle e|+\rho_g(t)|g\rangle\langle g|+\rho_s(t)|s\rangle\langle s|$ satisfies
\begin{equation}
    \begin{aligned}
        \frac{d}{dt}\rho_e(t) &=\kappa_p \rho_g(t)-\kappa_r \rho_e(t) -\kappa_{1} \rho_e(t)\\
        \frac{d}{dt}\rho_g(t) &=-\kappa_p \rho_g(t)+\kappa_r \rho_e(t)+\kappa_{2}\rho_s(t)\\
        \frac{d}{dt}\rho_s(t) &=\kappa_{1} \rho_e(t)-\kappa_2 \rho_s,
    \end{aligned}
    \label{SM:3lsRatesEqs}
\end{equation}
where the parameter $\kappa_1$ ($\kappa_2$) represents rates of transition between $|e\rangle$ and $|s\rangle$ ($|s\rangle$ and $|g\rangle$). Let us define $\sigma\equiv \kappa_2(\kappa_p+\kappa_r+\kappa_1)+\kappa_1\kappa_p$ and $K\equiv \kappa_p+\kappa_r+\kappa_1+\kappa_2$. Then, the solution for initial state $\rho(0)=|g\rangle \langle g|$ works out to be
\begin{equation}
    \begin{aligned}
     \rho_e(t)=\mu \left[1-(1+\nu)e^{-\lambda_1 t}-(1-\nu)e^{-\lambda_2 t}\right],
    \end{aligned}
    \label{rhoE}
\end{equation}
where
\begin{equation}
    \begin{aligned}
     \mu &=\frac{\kappa_2 \kappa_p}{\sigma}\\
     \nu &=\frac{\sigma+\kappa_1 \kappa_p-\kappa_2^2}{\kappa_2 \sqrt{K^2-4 \sigma}}\\
     \lambda_{1,2} &=\frac{1}{2}\left(K \pm \sqrt{K^2-4\sigma}\right),
    \end{aligned}
\end{equation}
which implies $\lambda_1+\lambda_2+\nu (\lambda_1-\lambda_2)>0$.
Note that although $\lambda_{1,2}$ gain an imaginary part when $K^2-4\sigma<0$, $\rho_e(t)$ in (\ref{rhoE}) remains real since this case yields $\rho_e(t)=\mu \left[1-(1+\nu)e^{-\lambda_1 t}\right] +c.c.$
The regression theorem allows us to determine
\begin{equation}
    \begin{aligned}
     \langle n \rangle &=\kappa_r \mu\\
     G^{(2)}(\tau) &=\kappa_r^2 \mu \rho_e(\tau),
    \end{aligned}
\end{equation}
which implies
\begin{equation}
\begin{aligned}
 P_1(w)& \approx \frac{T \kappa_r \mu}{2}  w\\
 P_{11}(w)& \approx \frac{(T\kappa_r \mu)^2}{4} \left[-\frac{\lambda_1 w-1+e^{-\lambda_1 w}}{\lambda_1^2}(1+\nu)\right.\\
 &\left.-\frac{\lambda_2 w-1+e^{-\lambda_2 w}}{\lambda_2^2}(1-\nu)+w^2\right].
\end{aligned}
\end{equation}
Mixing the light emitted from this three-level system with Poisson background noise gives rise to the correlation functions following as:
\begin{equation}
    \begin{aligned}
    \alpha(w) &\approx 1+\frac{f_3(w)-1}{(1+\bar{n}/\eta)^2}\\
    \beta(w) &\approx 1-\frac{\eta^2 w^2 T^2}{4} \left[1-f_3(w)\right]\\
    \gamma(w) &\approx 1+\frac{\eta w T}{2(1+\bar{n}/\eta)}\left[1-f_3(w)\right],
     \label{SM:corr3LSnoise}
    \end{aligned}
\end{equation}
where $\eta\equiv \kappa_r \mu$ and $f_3(w)$ gains
\begin{equation}
\begin{aligned}
 f_{3}(w)& \approx \frac{1}{4 w^2} \left[-\frac{\lambda_1 w-1+e^{-\lambda_1 w}}{\lambda_1^2}(1+\nu)\right.\\
 &\left.-\frac{\lambda_2 w-1+e^{-\lambda_2 w}}{\lambda_2^2}(1-\nu)+w^2\right].
\end{aligned}
\end{equation}
Because we can determine numerically $w_0$ such that $f_3(w_0)=1$, which implies the correlation functions reach unity, we can also establish numerically the unnormalized success $R\equiv P_{10}(w)w_0/w$.

\begin{figure*}[ht!]
\centerline {\includegraphics[width=0.9\linewidth]{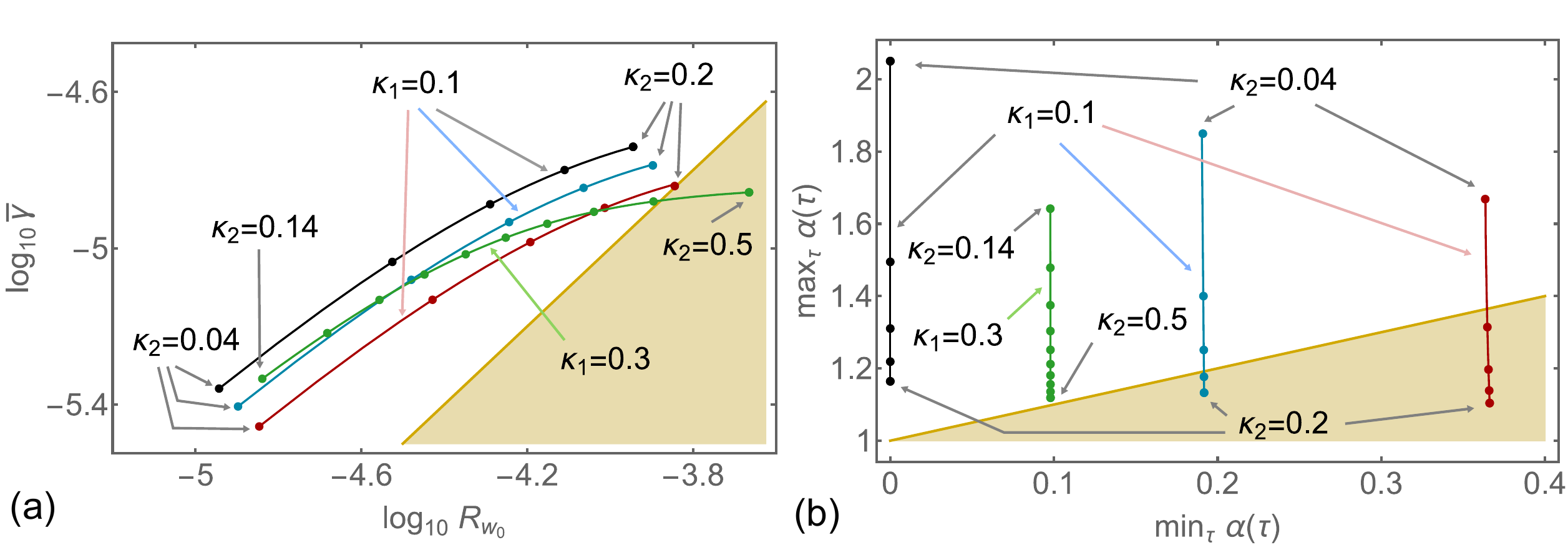}}
\caption{Simulation of light radiated from an emitter with a shelving state. We evaluate a photonic state employing either $\bar{\gamma}$ and $R_{w_0}$ in (\emph{a}) or $\max_{\tau}\alpha(\tau)$ and $\min_{\tau}\alpha(\tau)$ in (\emph{b}). Reaching the white region in (\emph{a}) or (\emph{b}) certifies presence of a shelving state in an emitter. In contrast, achieving only the yellow region in both (\emph{a}) and (\emph{b}) represents an inconclusive case when we neither certify nor confirm any aspect of an emitter. Here, we focus on the three-level dynamics determined by Eq.~(\ref{SM:3lsRatesEqs}) with parameters $\kappa_r=\kappa_p=1$ and various values of $\kappa_1$ (transition rate to the shelving state) and $\kappa_2$ (transition rate from the shelving state to the ground state). We assume a collection efficiency $T=10^{-4}$. The figure presents several examples of photonic states emerging from these dynamics. First, the black lines depict results of the dynamics that the parameters $\kappa_1=0.1$ and $\kappa_2 \in (0.04,0.2)$ specify. In this example, we assume no background noise deteriorates the emitted light. The black points correspond to $\kappa_2\in \{0.04,0.08,0.12,0.16,0.2\}$. Since the whole black lines are in the white regions, we certify the three-level dynamics for all $\kappa_2 \in (0.04,0.2)$ when using a criterion depicted either in plot (\emph{a}) or in (\emph{b}). Second, the blue and red line (blue and red point) in both plots show how  Poisson background noise affects the evaluated photonic state, represented by the black line (black points). We allow for Poisson background noise increasing the mean photon flux of the measured light by $10\%$ (blue) and $20\%$ (red). We certify the three-level dynamics for every $\kappa_2 \in (0.01,0.2)$ in both cases of the Poisson noise if we employ the criterion relying on $\bar{\gamma}$ and $R_{w_0}$, as depicted in plot (\emph{a}). In contrast, we fail in the certification based on $\max_{\tau}\alpha(\tau)$ and $\min_{\tau}\alpha(\tau)$, shown in plot (\emph{b}), if the emitter evolves with $\kappa_2 \in (0.15,0.2)$ and Poisson noise increases photon flux by $10\%$ or $\kappa_2 \in (0.07,0.2)$ and Poisson noise increases photon flux by $20\%$. Third, the green lines depict the position of a photonic state that results from the dynamics characterized by $\kappa_1=0.3$, $\kappa_2 \in \left(0.14,0.5\right)$ and Poissonian background noise causing growth in the mean photon flux by $10\%$. The green points show the cases when $\kappa_2 \in \{0.14,0.18,...,0.46,0.5\}$. Choosing $\kappa_2 \in \left(0.46,0.5\right)$ implies the certification in (\emph{b}) succeeds although these photonic states get in the inconclusive yellow region in (\emph{a}).}
\label{fig:compare}
\end{figure*}

\subsection*{Appendix C: Criteria} \label{SM:crit}
Here, we detail derivation of criteria rejecting models based only on emission from a two-level system. The notation we use for a general quantity $Q$ that characterizes the detection is either $Q\equiv Q(\rho,w)$ or $Q\equiv Q(w)$.
The former notation emphasizes that we simulate response of the detector on a photonic state $\rho$, i.e. $Q(\rho,w)$ corresponds to a function of parameters that determine the state $\rho$ and $w$ being the size of a time bin that we allow for in a simulation. The latter notation $Q(w)$ is used for referring to $Q$ that we can extract from time tag data achieved in an experimental realization, and therefore $Q$ depends only on $w$.

\subsubsection{Above-band pumping a two-level system with Poisson background noise}
The above-band pumping a two-level system  induces
\begin{equation}
\begin{aligned}
    &\alpha(\rho_{2LS},w,\tau) =\\
    &\frac{2\left[1-K (w-\tau)\right]-2e^{-K (w-\tau)}+K^2 (w-\tau)^2}{K^2 w^2}\\
    &+\frac{2e^{-K w}-e^{-K (w-\tau)}-e^{-K (w+\tau)}+K^2 \tau^2}{K^2 w^2},
\end{aligned}
    \label{SM:offres2LSAlpha}
\end{equation}
where we assume $\tau<w$ and $\rho_{2LS}$ represents a photonic state that the two-level system radiates and $K$ stands for a parameter that determines evolution of the two-level system. 
On the contrary, $\tau>w$ implies
\begin{equation}
\begin{aligned}
&\alpha(\rho_{2LS},w,\tau) =\\
&\frac{2e^{-K \tau}-e^{-K (\tau-w)}-e^{-K (w+\tau)}+K^2 \tau^2}{K^2 w^2}
\end{aligned}
\label{SM:offres2LSAlpha1}
\end{equation}

In both cases (\ref{SM:offres2LSAlpha}) and (\ref{SM:offres2LSAlpha1}), we gain $\alpha(\rho_{2LS},w)<1$, which guarantees $\beta(\rho_{2LS},w)<1$ and $\gamma(\rho_{2LS},w)>1$. It remains true even when the Poisson background noise deteriorates the state $\rho_{2LS}$.
However, general classical background noise $\rho_{\bar{n}}$ can in principle lead to $\alpha(\rho_{2LS}\otimes \rho_{\bar{n}},w)>1$. Here, we aim at criteria that reject the model states $\rho_m=\rho_{2LS}\otimes \rho_{\bar{n}}$ defined as a photonic state that emerges from mixing light emitted from the two-level system and classical background noise. First, we derive such a criterion employing the standard analysis of $\alpha(w,\tau)$ with variable delay time $\tau$ and fixed $w$. Second, we establish a criterion based on the correlation function $\gamma(w)$, which we evaluate for the delay time $\tau=0$, and demonstrate that these criteria are satisfied under different conditions.

\subsubsection{Correlation function $\alpha$}
To derive a criterion, we consider the correlation function $\alpha(w,\tau)$ in the limit $w\rightarrow 0$. Then, Eq.~(\ref{SM:rhoTLS}) with an assumption that classical background noise contributes independently to measured click events implies
\begin{equation}\label{SM:alphaT}
\begin{aligned}
    \alpha(\rho_m,\tau)=\frac{\mu^2 \left[1-e^{-K \tau}\right]+2 \mu \bar{n}+g^{(2)}(\tau)\bar{n}^2}{(\mu+\bar{n})^2},
\end{aligned}
\end{equation}
where $\mu$ and $K$ characterize dynamics of the two-level system, $\bar{n}$ is the mean number of noisy photons and $g^{(2)}(\tau)$ is the second-order correlation function of the classical background noise. Further, we define
\begin{equation}\label{SM:defD}
d(\rho_m)\equiv \sup_{\tau}\alpha(\rho_m,\tau)-\inf_{\tau}\alpha(\rho_m,\tau).
\end{equation}
and derive a criterion from maximizing $d(\rho_m)$ over the parameters $\mu$, $K$, $\bar{n}$ and any function $g^{(2)}(\tau) \geq 1$ that monotonously drops with $\tau$ \cite{Glauber1963}. Since the maximum over such functions $g^{(2)}(\tau)$ occurs when $g^{(2)}(\tau)$ is a constant function, we can find out $\max_{\rho_m}d(\rho_m)=1$. Thus, measuring
\begin{equation}\label{SM:critA}
\sup_{\tau}\alpha(\tau)-\inf_{\tau}\alpha(\tau)>1
\end{equation}
rejects all the situations reachable by the model states $\rho_m$ in Eq.~(\ref{SM:alphaT}).

\begin{figure*}[ht]
\centerline {\includegraphics[width=0.9 \linewidth]{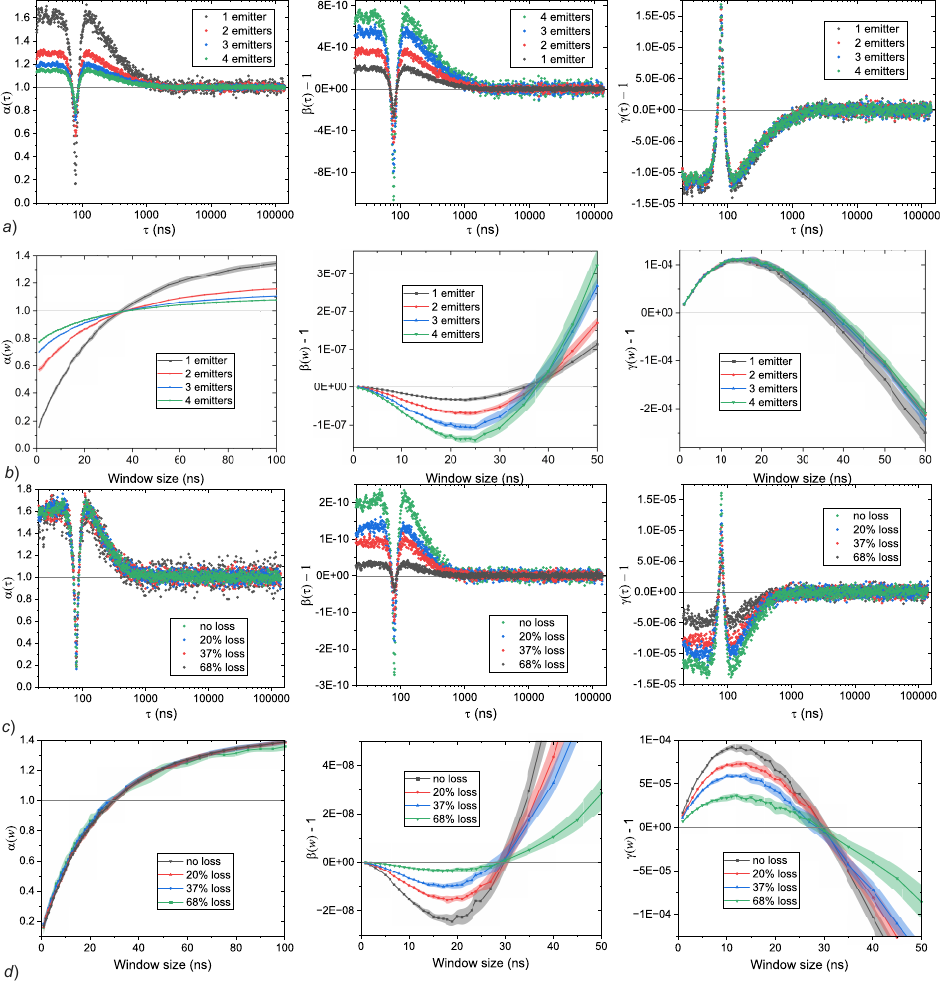}}
\caption{\emph{a}) Plots of the correlation functions $\alpha(w,\tau)$, $\beta(w,\tau)$, and $\gamma(w,\tau)$ evaluated from the experimental data with a time-bin size of $w=1$~ns and varying number of equally contributing quantum emitters. Here the time axis is shifted by 80~ns for better illustration of the plotted functions. The horizontal line where the function value is equal to unity shows the classicality threshold: Values of the correlation functions $\alpha$, $\beta$ ($\gamma$) below (above) this line correspond to non-classical light. Note that the plots for $\gamma(\tau)$ overlap, demonstrating invariance of $\gamma(w,\tau)$ over the number of contributing quantum emitters. \emph{b}) Plots of the correlation functions $\alpha(w,0)$, $\beta(w,0)$, and $\gamma(w,0)$ evaluated from the same experimental data as in \emph{a}) but with variable time-bin size $w$. Data points are connected for clarity, and shaded regions show the error range. The horizontal line at $\alpha(w)=\beta(w)=\gamma(w)=1$ shows the classicality threshold. Note that the plots for $\gamma(w)$ overlap, demonstrating invariance of $\gamma(w,\tau)$ over the number of contributing quantum emitters.
\emph{c}) Plots of the correlation functions $\alpha(w,\tau)$, $\beta(w,\tau)$, and $\gamma(w,\tau)$ evaluated from the experimental data with a time-bin size of $w=1$~ns and varying level of loss in the detection. Here the time axis is shifted by 80~ns for better illustration of the plotted functions. The horizontal line where the function value is equal to unity shows the classicality threshold: Values of the correlation functions $\alpha$, $\beta$ ($\gamma$) below (above) this line correspond to non-classical light. Note that the plots for $\alpha(\tau)$ overlap, demonstrating invariance of $\alpha(w,\tau)$ over loss level. \emph{d}) Plots of the correlation functions $\alpha(w,0)$, $\beta(w,0)$, and $\gamma(w,0)$ evaluated from the experimental data. Data points are connected for clarity, and shaded regions show the error range. The horizontal line at $\alpha(w)=\beta(w)=\gamma(w)=1$ shows the classicality threshold. Note that the plots for $\alpha(w)$ overlap, demonstrating invariance of $\alpha(w,\tau)$ over loss level.}
\label{fig:abg_tau_merge}
\end{figure*}

\subsubsection{Correlation function $\gamma$}\label{SM:parGamma}
Here, we focus on the criterion exploiting the correlation function $\gamma(w) \equiv P_{10}(w)/P_0(w)/P_1(w)$. Using the regression theorem \cite{Loudon2000}, we express the moments $\widetilde{n}(\rho_m,w)$ and $\widetilde{g}^{(2)}(\rho_m,w)$ defined in Eq.~(\ref{SM:moments}) following as:
\begin{equation}
\begin{aligned}
    \widetilde{g}^{(2)}(\rho_m,w)& = \frac{1}{K w^2\left(\eta+\bar{n}\right)^2}\left\{\eta^2\left[2(1-K w)-2e^{-K w}\right.\right.\\
    &\left.\left. K^2 w^2\right]+Kw^2\left[2 \eta \bar{n}_{bn}+\widetilde{g}^{(2)}_{bn}(w)\bar{n}^2_{bn}\right]\right\}\\
    \widetilde{n}(\rho_m,w)&=T(\eta+\bar{n}_{bn})w,
\end{aligned}
\end{equation}
where $\bar{n}_{bn} w$ is a mean photon number of the background noise, $\widetilde{g}^{(2)}_{bn}(w)$ corresponds to the second moment, defined in Eq.~(\ref{SM:noiseG2w}), that characterizes the background noise and, finally, $\eta$ and $K$ determine evolution of a two-level system.
The correlation function $\gamma(\rho_m,w)$ works out to be
\begin{equation}
    \gamma(\rho_m,w)\approx 1+T \bar{n}(\rho_m,w)\left[1-\widetilde{g}^{(2)}(\rho_m,w)\right].
\end{equation}
Thus, the classical background noise results in $\gamma(\rho_m,w)=1$ for a given $w$ if
\begin{equation}\label{SM:w0Eq}
    \bar{n}_{bn}^2\left[\widetilde{g}_{bn}(w)-w^2\right]=\frac{2\eta^2}{K^2 w^2}\left(K w-1+e^{-K w}\right).
\end{equation}
Excluding $\bar{n}_{bn}$ from (\ref{SM:w0Eq}) yields
\begin{equation}
\begin{aligned}
 &\gamma(\rho_m,w)=1+\frac{\eta T}{K w_0\left[K+\sqrt{\frac{2f(K w)}{\widetilde{g}_{w_0}}}\right]}\\
 &\times \left(w_0 f(K w)-\frac{\widetilde{g}_{w}}{\widetilde{g}_{w_0}}w f(K w_0)\right),
\end{aligned}
\label{SM:gamma}
\end{equation}
where $f(x)\equiv x-1+\exp(-x)$ and $\widetilde{g}_{w}\equiv \left[\widetilde{g}_{bn}(w)-2 w^2\right]/w^2$ and $w_0$ stands for the size of a time bin that satisfies (\ref{SM:w0Eq}).

Furthermore, we employ the parameter $R_w(\rho_m)$ defined in Eq.~(\ref{SM:Rpar}) as
\begin{equation}
\begin{aligned}
    R_{w_0}(\rho_m)\equiv \frac{P_{01}(\rho_m)w_0}{w}
\end{aligned}
\end{equation}
with $w_0$ leading to $\gamma(w_0)=1$
and approximate its value for the model state $\rho_m$ following as: 
\begin{equation}
\begin{aligned}
 R(\rho_m)\approx T \eta\left[\frac{w_0}{2}+\frac{1}{K}\sqrt{\frac{2 f(K w_0)}{\widetilde{g}_{w_0}}}\right],
\end{aligned}
\label{SM:Rw}
\end{equation}
which is independent of $w$. To identify a criterion, we introduce the function
\begin{equation}
    F_{w_0}(w,K,\eta,\widetilde{g}_{w},\widetilde{g}_{w_0})\equiv \frac{\gamma(\rho_m,w)-1}{R_{w_0}(\rho_m)},
\end{equation}
which is independent of the detection efficiency $T$, and calculate its supremum
\begin{equation}
    F_{w_0}=\sup_{w,K,\eta,\widetilde{g}_{w},\widetilde{g}_{w_0}} F_{w_0}(K,\eta,\widetilde{g}_{w},\widetilde{g}_{w_0}).
\end{equation}
We consider the following to obtain $F_{w_0}$. First, according to Eq.~(\ref{SM:gamma}) and the constraint $\widetilde{g}_{w}\leq \widetilde{g}_{w_0}$ (satisfied for any classical noise), we conclude that the maximum $F_{w,0}$ occurs for $\widetilde{g}_{w}=\widetilde{g}_{w_0}$, which implies
\begin{equation}\label{SM:fmax}
    F_{w_0}=\sup_{w,K,\widetilde{g}_{w_0}} \frac{2 w\left[f(K w)w_0^2-f(K w_0)w^2\right]}{\left[K w_0+\sqrt{f(K w_0)/\widetilde{g}_{w_0}}\right]^2}.
\end{equation}
Second, Eq.~(\ref{SM:fmax}) indicates that the supremum $F_{w_0}$ happens in the limit $\widetilde{g}_{w_0}\rightarrow \infty$, which leads to
\begin{equation}\label{SM:supG}
    F_{w_0}=\sup_{w,K} G_{w_0}(w,K)
\end{equation}
with $G_{w_0}(w,K)\equiv w\left[f(K w)w_0^2-f(K w_0)w^2\right]/w_0^2/K$. Finally, we calculate numerically the supremum in Eq.~(\ref{SM:supG}), which leads to $F_{w_0}=0.093$,
independently of $w_0$. To show that $F_{w_0}$ is a constant function of $w_0$, we employ the definition of $G(w,w_0,K)$ to prove the identity
\begin{equation}\label{SM:derGEq}
\begin{aligned}
    w_0 \partial_{w_0} G_{w_0}(w,K)&=K \partial_K G_{w_0}(w,K)\\
    &-w \partial_{w} G_{w_0}(w,K).
\end{aligned}
\end{equation}
Because the supremum $F_{w_0}$ happens for a given $w_0$ when $\nabla G_{w_0}(w,K)=0$, Eq.~(\ref{SM:derGEq}) guarantees that $\partial_{w_0}F_{w_0}=0$, and therefore $F_{w_0}$ is independent of $w_0$. 

The supremum $F_{w_0}$ allows us to formulate a criterion. A measurement achieving
\begin{equation}\label{SM:critG}
\gamma(w)-1>F_{w_0} R_{w_0}
\end{equation}
for some $w$ can not be explained as detection of any model states $\rho_m$. To compare the criterion in Eq.~(\ref{SM:critG}) with the criterion in Eq.~(\ref{SM:critA}), we allow for an emitter with the shelving state that undergoes particular evolution given by Eq.~(\ref{SM:3lsRatesEqs}) with $\kappa_r=\kappa_p=1$, $\kappa_1=0.1$ and $\kappa_2 \in (0.1,0.25)$. In an ideal case, when no background Poisson noise deteriorates the emitted light, both criteria in Eqs.~(\ref{SM:critA}) and (\ref{SM:critG}) get satisfied for any $\kappa_2 \in (0.1,0.25)$. However, even small amount of Poisson noise contributing to the measurement causes that the certification based on the criterion with $\alpha(\tau)$ fails, as depicted in Fig.~\ref{fig:compare}. On the contrary, the criterion relying on $\gamma(w)$ is less sensitive to the Poisson background noise in this example of the emitted photonic state.

Furthermore, we demonstrate an analogical approach to derive a criterion based on $\gamma(w_{0})$ and the parameter $R_{w_0}(\rho_m)$ with $w_0$ maximizing $\gamma(w)$, i.e. $\gamma(w_0)=\max_w \gamma(w)$. Evaluating $P_{10}(w)$ in Eq.~(\ref{SM:Rpar}) implies the identity
\begin{equation}
R_{w_0}(\rho_m)=P_{10}(\rho_m),
\end{equation}
in which case $R_{w_0}$ corresponds to the probability $P_{10}(\rho_m)$, measured in a well-defined temporal mode. For a given model state $\rho_m$, we determine $w_{0}$ from the equation
\begin{equation}\label{SM:derGAlt}
    \partial_w \gamma(\rho_{m},w)=0,
\end{equation}
which leads to
\begin{equation}\label{SM:idNBN}
\bar{n}^2_{bn}\left(\widetilde{g}^{(2)}(w)-w^2\right)K^4=2 \kappa_p \kappa_r \left[1-e^{-K w}\left(1+K w\right)\right].
\end{equation}
In analogy to the procedure yielding the criterion in Eq.~(\ref{SM:critG}), we determine $\bar{n}_{bn}$
implying that Eq.~(\ref{SM:idNBN}) gets fulfilled at a given $w$. Substituting such $\bar{n}_{bn}$ into the ratio
\begin{equation}
E_{w_0}(\rho_m)=\frac{\gamma(\rho_m,w_0)}{R_{w_0}(\rho_m)}
\end{equation}
allows us to derive the criterion
\begin{equation}\label{SM:critAlt}
\gamma(w_0)-1>E_{w_0} P_{10}(w_0),
\end{equation}
where $E_{w_0}=\max_{\rho_m}E_{w_0}(\rho_m)$ gains $E_{w_0}\approx 0.27$ independently of $w_0$.
Compared to the criterion in Eq.~(\ref{SM:critG}), the criterion in Eq.~(\ref{SM:critAlt}) relies on different normalization in $R_{w_0}$, and therefore these two criteria are not equivalent.

\subsection*{Appendix D: Correlation functions for specific cases}\label{sec_extraplots}



Here we present plots of correlation functions $\alpha(w,\tau)$, $\beta(w,\tau)$, and $\gamma(w,\tau)$ for cases of varying number of (equally contributing) quantum emitters and varying level of loss.



\subsubsection{Varying number of quantum emitters}

When one deals with solid-state quantum emitters, it is experimentally challenging to find samples with well-defined number of equally bright and uncorrelated single-photon emitters. In order to mimic multiple independent quantum emitters, we record photocounts from the same single quantum emitter in several experimental runs, and then merge photocounts from two, three, and four data sets to represent two, three, and four quantum emitters, respectively. This guarantees that the emitters have the same internal properties (brightness and auto-correlation) and are uncorrelated at the same time, since the consecutive experimental runs are independent. The corresponding correlation functions are shown in Figs.~\ref{fig:abg_tau_merge} \emph{a}) and \emph{b}).

\subsubsection{Varying loss}

In order to record data with varying loss level, we subsequently introduce neutral density filters of optical density 0.1, 0.2, and 0.5 in the detection part of the setup. The corresponding correlation functions are shown in Figs.~\ref{fig:abg_tau_merge} \emph{c}) and \emph{d}).

\bibliography{modelBib.bib}

\end{document}